\documentclass{PoS}
\title{Energy, Rapidity and Transverse Momentum Dependence of Multiplicity Fluctuations in Heavy Ion Collisions at CERN SPS }

\ShortTitle{Energy Dependence of Multiplicity Fluctuations at CERN SPS}

\author{\speaker{Benjamin Lungwitz}\\
        Institut für Kernphysik, Universität Frankfurt, Germany\\
        E-mail: \email{lungwitz@ikf.uni-frankfurt.de}}

\author{for the NA49 Collaboration:}

\author{
C.~Alt$^{9}$, T.~Anticic$^{23}$, B.~Baatar$^{8}$,D.~Barna$^{4}$,
J.~Bartke$^{6}$, L.~Betev$^{10}$, H.~Bia{\l}\-kowska$^{20}$,
C.~Blume$^{9}$,  B.~Boimska$^{20}$, M.~Botje$^{1}$
% ...
\\

\vspace{0.5cm}
$^{1}$NIKHEF, Amsterdam, Netherlands. \\
% ...
$^{4}$KFKI Research Institute for Particle and Nuclear Physics, Budapest, Hungary.\\
% ...
$^{6}$Henryk Niewodniczanski Institute of Nuclear Physics, Polish Academy of Sciences, Cracow, Poland.\\
% ...
$^{8}$Joint Institute for Nuclear Research, Dubna, Russia.\\
$^{9}$Fachbereich Physik der Universit\"{a}t, Frankfurt, Germany.\\
$^{10}$CERN, Geneva, Switzerland.\\
% ...
$^{20}$Institute for Nuclear Studies, Warsaw, Poland.\\
% ...
$^{23}$Rudjer Boskovic Institute, Zagreb, Croatia.\\
% ...
}

\abstract{Multiplicity fluctuations in the forward hemisphere were studied for positively, 
negatively and all charged hadrons produced
in central Pb+Pb collisions at $20A$, $30A$, $40A$, $80A$ and $158A$ GeV. 
The multiplicity distributions and their scaled variances are presented as a
function of collision energy, rapidity and transverse momentum.
The distributions have a bell-like shape and  the scaled variance
changes monotonously with energy in the range from 0.8 to 1.2.  
No indication of the critical point is observed.

The string-hadronic model UrQMD reproduces results on the scaled variance.
The predictions of the hadron-resonance gas model obtained within the grand-canonical 
and canonical ensembles for the scaled variance disagree with the data.
}

\FullConference{Critical Point and Onset of Deconfinement
          4th International Workshop\\
		 July 9-13 2007\\
		 GSI Darmstadt,Germany}

\begin{document}

\section{Introduction}

At high energy densities ($\approx 1\, GeV/fm^3$) a phase transition between hadron gas and quark-gluon-plasma (QGP) is expected to occur.
There are indications that at RHIC and top SPS energies the quark-gluon-plasma is created at the early stage of heavy ion 
collisions~\cite{Heinz:2000bk, Gyulassy:2004zy}.
The energy dependence of 
various observables shows anomalies at low SPS energies~\cite{Afanasiev:2002mx}
which might be related to the onset of deconfinement~\cite{Gazdzicki:1998vd}. 
It is predicted \cite{Gazdzicki:2003bb} that the onset of deconfinement should lead 
to a non-monotonous behaviour in the relative fluctuations of entropy to energy, which are related to the fluctuations in
multiplicity.

 multiplicity fluctuations, the so-called "shark fin".
Lattice QCD calculations suggest furthermore the existence of a critical point in the phase diagram of strongly interacting matter which separates the 
first order phase boundary at high baryo-chemical potentials and low temperature from a crossover at low baryo-chemical
potential and high temperature. 
An increase of multiplicity fluctuations near the critical point of strongly interacting matter is expected~\cite{Stephanov:1999zu}.

In statistical models the widths of the multiplicity distributions are dependent on the conservation laws the system obeys.
Even though the mean multiplicity is the same in the infinite volume limit 
for different statistical ensembles, this is not true for
multiplicity fluctuations~\cite{Begun:2004gs}.
Therefore multiplicity fluctuations in nuclear collisions provide a unique tool 
for testing the influence of conservation laws in relativistic gases.

The predictions for multiplicity fluctuations obtained by the HSD~\cite{Konchakovski:2005hq}
and UrQMD~\cite{Lungwitz:2007uc} models are different from the hadron-gas predictions. 
Therefore the multiplicity fluctuations might allow to distinguish between these models.

Results on the centrality dependence of multiplicity fluctuations in Pb+Pb collisions
obtained by the NA49~\cite{Alt:2006jr} and WA98~\cite{Aggarwal:2001aa}
collaborations at top SPS energies show an increase
of multiplicity fluctuations with decreasing centrality of the collision for forward rapidities.
This increase might be interpreted as fluctuations in the number of target participants, 
which contribute to the projectile hemisphere~\cite{Gazdzicki:2005rr} or
as an effect of correlations between produced particles~\cite{Rybczynski:2004zi}.
A similar increase of multiplicity fluctuations is observed at midrapidity by the 
PHENIX~\cite{Mitchell:2005at,Homma:2007qh} collaboration at RHIC energies.

In this report the energy, rapidity and transverse momentum dependence of multiplicity fluctuations in very central Pb+Pb collisions
measured by the NA49 experiment is presented and compared to predictions of the UrQMD and a hadron-resonance gas model~\cite{Begun:2006uu}.

\section{Measure of Multiplicity Fluctuations}\label{c_measure}

Let $P(n)$ denote the probability
to observe a particle multiplicity $n$ ($\sum_n P(n) = 1$)
in a high energy nuclear collision. 

The scaled variance $\omega$, used in this paper as a measure 
of multiplicity fluctuations, is defined as
\begin{equation}
\omega=\frac{Var(n)}{<n>}=\frac{<n^2>-<n>^2}{<n>},
\end{equation}
where $Var(n)=\sum_n (n-<n>)^2 P(n)$ and $<n>=\sum n \cdot P(n)$ 
are variance and mean of the multiplicity distribution, respectively.

In many models this measure is independent of the number of particle production sources.
First, in grand-canonical statistical models  neglecting quantum effects and resonance 
decays the multiplicity distribution is a Poisson one.
The variance of a Poisson distribution is equal to its mean, 
and thus the scaled variance is $\omega=1$, independently of mean multiplicity.
Second, in the Wounded Nucleon Model~\cite{Bialas:1976ed}, 
the scaled variance in $A+A$ collisions is the same as in nucleon-nucleon collisions provided 
the number of wounded nucleons is fixed.

If the particles are not correlated in momentum space, 
the scaled variance in a limited acceptance is related to the scaled variance 
in full phase-space ("4$\pi$") as~\cite{Begun:2006uu}:  
\begin{equation}\label{wscale}
\omega_{acc}=p \cdot (\omega_{4\pi} -1) +1,
\end{equation}
where $p$ denotes the mean fraction of particles measured in the corresponding acceptance.
Note that effects like resonance decays, 
quantum statistics and energy-momentum conservation introduce correlations in momentum space. 
Therefore the scaling described by equation \ref{wscale} is in general violated.

\section{The NA49 Experiment}\label{c_NA49}

The NA49 detector~\cite{Afanasev:1999iu}
is a large acceptance fixed target hadron spectrometer. Its main devices are four large 
volume time projection chambers (TPCs).
Two of them, called vertex TPCs (VTPC1 and 2), are located in two superconducting dipole magnets with a total bending power up to $7.8$ Tm. 
The magnetic field used at $158A$ GeV was scaled down in proportion to the beam energy for lower energies.
The other two TPCs (MTPC-L and MTPC-R) are installed behind the magnets on the left and the
right side of the beam line allowing precise particle tracking in the high density region of heavy ion collisions.
The measurement of the energy loss $dE/dx$ in the detector gas allows particle identification in a large momentum range. It is complemented 
by time of flight (TOF) detectors measuring particles at mid-rapidity.
In this analysis $dE/dx$ information is used only to reject electrons.

The target is located $80$ cm in front of the first vertex TPC. 
Three beam-position-detectors (BPDs) allow a precise determination of the 
point where the beam hits the target foil. 
The centrality of a collision is determined by measuring 
the forward going energy of projectile spectators
in the veto calorimeter (VCAL, see section~\ref{centsel}). 
The acceptance of the veto calorimeter is adjusted at 
each energy by a proper setup of the collimator.

\subsection{Event Selection}
\label{ev_sel}

In order to get a "clean" sample of events excluding for instance 
collisions outside the target or event pileup,
the fit of the interaction point, based on the reconstructed tracks, has to be successful and has to be close to
the position obtained by extrapolation
of the beam particle measured by the beam position detectors to the target foil.

The event cuts have a small influence on the scaled variance, 
the results differ by less than $1\%$ when only the cut requirement of a successful
fit of the main vertex is used.

\subsection{Centrality Selection}\label{centsel}

Fluctuations in the number of participants lead to an increase of multiplicity fluctuations. 
In a superposition model the total multiplicity is the sum of the number of particles 
produced by different independent particle production sources.
In this model the scaled variance has two contributions. The first is due to the fluctuations of the number of 
particles emitted by a single source, 
the second is due to the fluctuations in the number of sources.
In order to minimize the latter
the centrality variation in the ensemble of events should be as small as possible.

The downstream veto calorimeter~\cite{DeMarzo:1983gd} of NA49, originally designed for NA5,  allows a determination of the energy in the 
projectile spectator region~\cite{Appelshauser:1998tt}.
A collimator in front of the calorimeter is located $25$~m downstream from the target and is adjusted for each energy in such a way that all projectile 
spectator protons, neutrons and fragments can reach the veto calorimeter. 
Acceptance tables for the veto calorimeter in $p$, $p_T$ and
$\phi$ can be obtained on the author`s website~\cite{accTables}.

The projectile centrality $C_{Proj}$ of an event with a veto energy $E_{Veto}$ can be calculated using the known trigger centrality 
$C_{trig}=\frac{\sigma_{trig}}{\sigma_{inel}}$ and the veto energy distribution as:
\begin{equation}
C_{Proj}=\frac{\sigma_{E_{veto}}}{\sigma_{inel}}=C_{trig} \cdot \frac{\int_{0}^{E_{veto}}{dN/dE_{veto,trig}} }{\int_{0}^{\infty} {dN/dE_{veto,trig}}},\label{ctrig_eq}
\end{equation}
where $dN/dE_{veto,trig}$ is the veto energy distribution for a given trigger.

The resolution of the veto calorimeter was estimated in~\cite{Alt:2006jr}. 
In order to check this parametrization, the distribution of the spectators were simulated by the SHIELD model~\cite{Dementev:1997ca}. 
A simulation performed at $20A$ and $158A$ GeV
including the geometry of the NA49 detector and the non-uniformity of the veto calorimeter confirms the parametrization
as an upper limit.
The corresponding fluctuations are expected to increase the scaled variance in the data by less than $1\%$; no correction (which would decrease 
the scaled variance) is
applied to the data.

The veto calorimeter response can in principle change with time (aging effects, etc.). Therefore a time dependent calibration of the veto energy 
was applied.
However, the effect of this calibration on the scaled variance turned out to be very small ($<1\%$).

\begin{figure}
\begin{center}
\includegraphics[height=5.5cm]{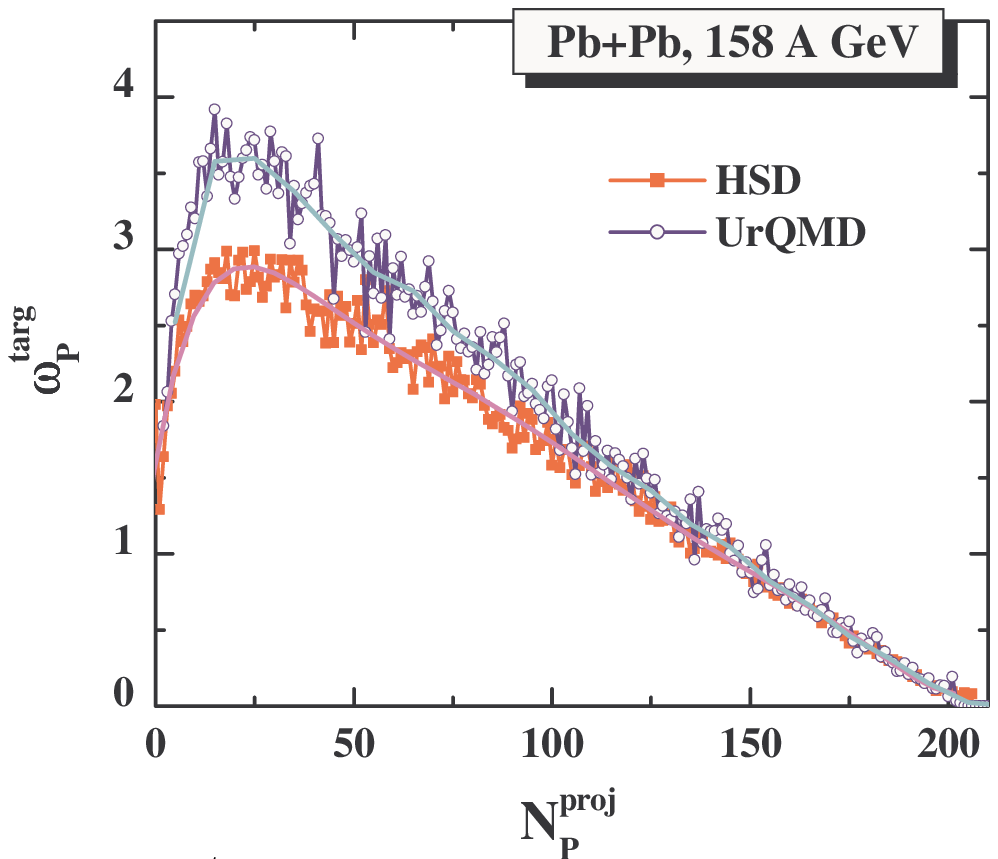}
\includegraphics[height=5.5cm]{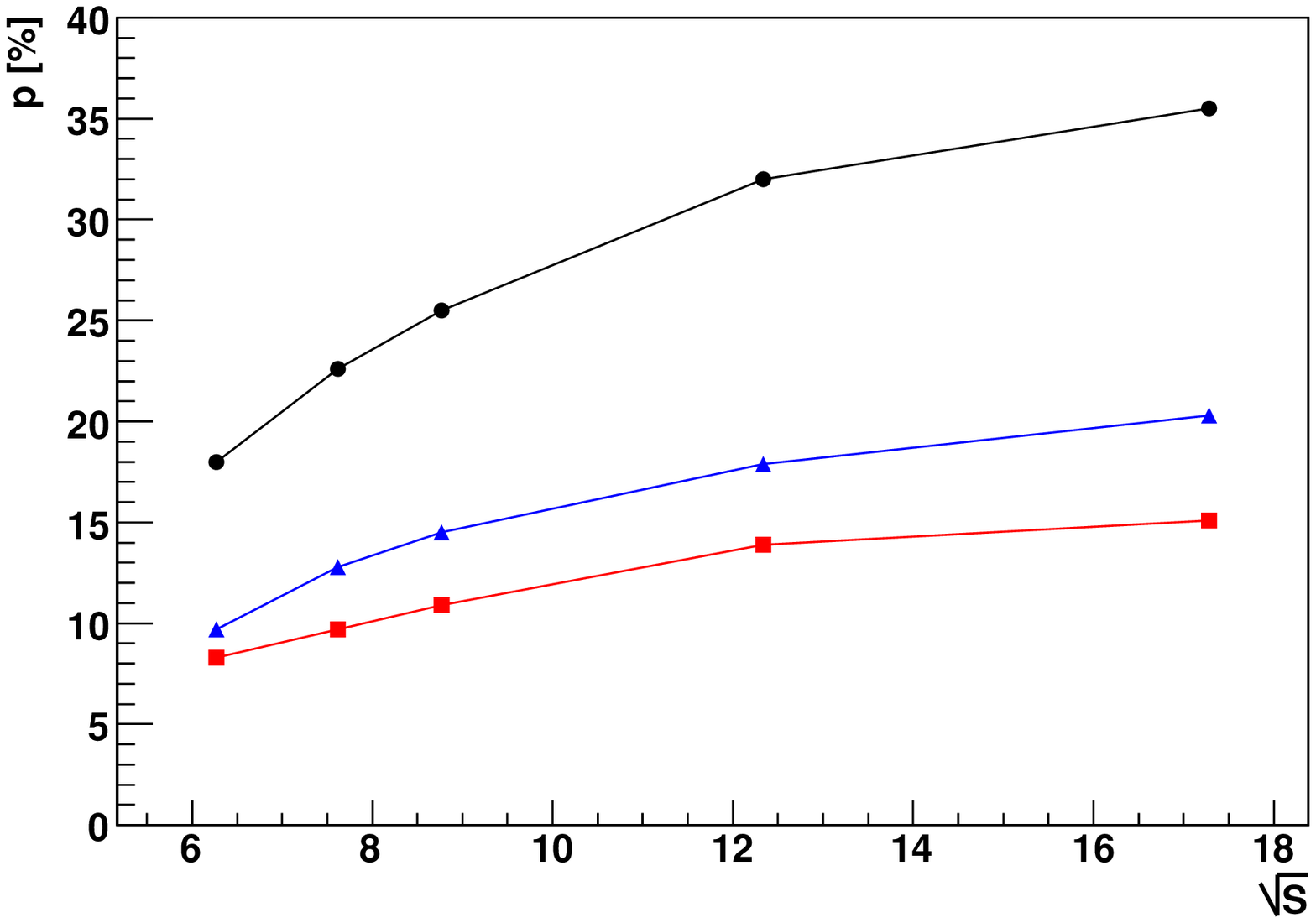}
\end{center}
\caption{\label{nptarg_fluct}\label{accept_ed}
Left: Fluctuations in the number of target participants for a fixed number of projectile participants in the UrQMD and HSD~\cite{Konchakovski:2005hq}
models.
Right: Fraction of total mean multiplicity in the acceptance. 
Black circles: $0<y(\pi)<y_{beam}$, red boxes: $0<y(\pi)<1$, blue triangles: $1<y(\pi)<y_{beam}$.}
\end{figure}

When fixing the projectile centrality $C_{Proj}$ by equation~\ref{ctrig_eq}, which fixes the number of projectile participants $N_P^{Proj}$, 
the number of target participants
$N_P^{Targ}$ can still fluctuate. 
This means that the total number of participants is not rigorously constant which may induce residual fluctuations.
The fluctuations of the target participants obtained by UrQMD and HSD simulations~\cite{Konchakovski:2005hq},
expressed as their scaled variance $\omega_P^{Targ}=Var(N_P^{Targ})/<N_P^{Targ}>$, are shown in figure~\ref{nptarg_fluct}.
For non-central collisions the target participants strongly fluctuate, even for a fixed number of projectile participants.
In the experimentally observed centrality dependence of multiplicity fluctuations~\cite{Alt:2006jr}, an increase of scaled variance in the 
forward hemisphere 
with decreasing centrality is detected, which might be related to the increasing target participant fluctuations~\cite{Gazdzicki:2005rr}. 

For further analysis, the $1\%$ most central collisions (according to their veto energy) are selected in order to minimize the flucutations 
in the number of participants. 
For these very central collisions, the fluctuations in the number of target participants are expected to be smallest and its scaled variance
$\omega_P^{Targ}$ is expected to be about $0.1$.

In order to study the influence of target participant fluctuations and
non-spectator particles in the veto calorimeter, the energy dependence of the scaled variance is calculated in the UrQMD 1.3 model both for 
collisions with zero
impact parameter and for collisions selected according to their veto energy. The difference in the scaled variance in the forward acceptance is 
smaller than $2\%$ for 
negative hadrons, smaller than $3\%$ for positive and smaller than $4\%$ for all charged hadrons.
In the midrapidity region the influence of the fluctuations of target participants to the scaled variance is expected to be much larger. 
In the UrQMD model the scaled variances are up to $6\%$ for negative, $9\%$ for positive and up to $13\%$ for all charged 
hadrons larger 
when selecting events by their veto energy than the
corresponding values for a zero impact parameter of the collision.

In order to check the influence of the centrality selection, in addition the scaled variance for the $0.5\%$ most central collisions was determined. 
The difference
to the values obtained for the $1\%$ most central collisions is smaller than $3\%$ for positive, $2\%$ for negative and $5\%$ for all charged hadrons.

\subsection{Track Selection}\label{trsel}

Since detector effects like track reconstruction efficiency might have a significant influence on multiplicity fluctuations, it is important to select a 
very clean track sample for the analysis.

Reconstruction inefficiencies mostly occur for tracks with a very low number of points in the time projection
chambers (TPCs) or for tracks which only have points in the first vertex TPC (VTPC1) or in the main TPC. These tracks are not used for this analysis.
Acceptance tables in $y(\pi)$, $p_T$ and $\phi$ can be obtained at the author`s website~\cite{accTables}. 
Only tracks in the rapidity interval starting at midrapidity and ending at beam rapidity are taken. 

In order to study the multiplicity fluctuations differentially, the total rapidity interval $0<y(\pi)<y_{beam}$ is divided into two parts,
the "midrapidity" ($0<y(\pi)<1$) and the "forward" ($1<y(\pi)<y_{beam}$) region. The fraction of $4\pi$ multiplicity which is accepted in the different
acceptance intervals is shown in figure~\ref{accept_ed}. 
Note that the acceptance used for this analysis is larger than the one used for the 
preliminary data shown in \cite{Lungwitz:2006cx,Lungwitz:2006cy}.

In order to decrease the contribution of weakly decaying particles, the measured tracks are extrapolated back to the target plane. The distance of the point, where
the extrapolated track hits the target plane, to the main vertex of the collision is called track impact parameter. All tracks with too high track impact parameters
are rejected.
For removing the electron contribution, tracks with a too high energy loss in the detector gas are rejected.
The influence of both cuts on the resulting scaled variance is small (see section~\ref{syserr}).

\subsection{Systematic Errors}\label{syserr}

The influence of several effects on the scaled variance have been studied.
These include the event selection criteria, the resolution and the calibration of the veto calorimeter and the track selection criteria.
The total systematic error is estimated as the maximum of these effects. It is $2\%$ for positively 
and negatively charged hadrons and $3\%$ for all charged hadrons.

In order to estimate the effect of centrality selection, the $0.5\%$ most central collisions are selected. The scaled variance for this stricter 
selection
is up to $5\%$ different from the scaled variance obtained for the $1\%$ most central collisions. As the centrality selection is a well-defined 
procedure
and can be repeated in model calculations, the difference of the $0.5\%$ and $1\%$ most central collisions is not part of the systematic error.

In the midrapidity region at top SPS energy ($158A$ GeV) a high track density causes track losses in events with a high multiplicity. 
This effect can be seen as an asymmetry in the ratio of the measured multiplicity distribution to the Poisson distribution.
Therefore the results on the scaled variance in the midrapidity and full experimental 
acceptance at $158A$ GeV are not shown in this paper.

\section{Results on Multiplicity Fluctuations}\label{mult_res}

In this chapter the results on the multiplicity fluctuations for negatively, positively and all charged hadrons
are presented for Pb+Pb collisions at $20A$, $40A$, $80A$ and $158A$ GeV.
In order to minimize the fluctuations in the number of participants, the $1\%$ most central collisions according to the energy of projectile spectators 
measured in the veto calorimeter are selected (see section~\ref{centsel}).
For a more differential study, the phase-space is divided into three different rapidity intervals: $0<y(\pi)<y_{beam}$,
$0<y(\pi)<1$ and $1<y(\pi)<y_{beam}$ (see section~\ref{trsel}).

%\subsection{Multiplicity Distributions}

\begin{figure}
\begin{center}
\includegraphics[height=5cm]{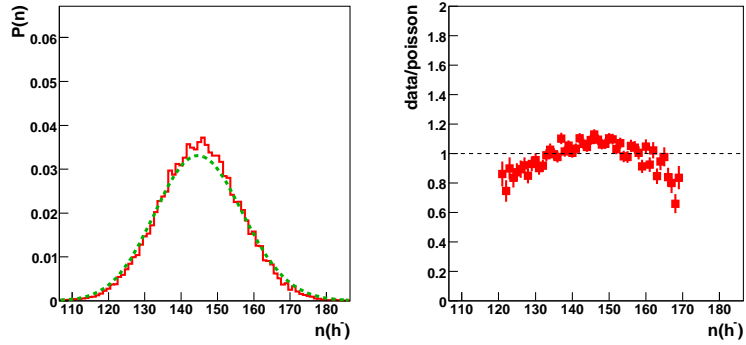}
\end{center}
\caption{\label{mult_dist}Left: Multiplicity distribution of experimental data in comparison to a corresponding Poisson
distribution for negatively charged hadrons in Pb+Pb collisions at $158A$ GeV in the forward acceptance. Right: Ratio of the measured distribution
over a Poisson distribution.}
\end{figure}

As an example, the multiplicity distribution in the forward acceptance at $158A$ GeV is shown in figure~\ref{mult_dist}.

The multiplicity distributions all have a bell-like shape, 
no significant amount of events with a very high or very low multiplicity are observed.
They are compared to a Poisson distribution with the same mean multiplicity.

The measured multiplicity distributions are significantly narrower than the Poisson ones 
in the forward acceptance for positively and negatively charged hadrons at all energies. In the midrapidity acceptance the measured distributions are
wider than the Poisson ones. 
The distributions for all charged hadrons are broader than the ones for a single charge.

%\subsection{Energy Dependence of Scaled Variance}

\begin{figure}
\includegraphics[height=3.5cm]{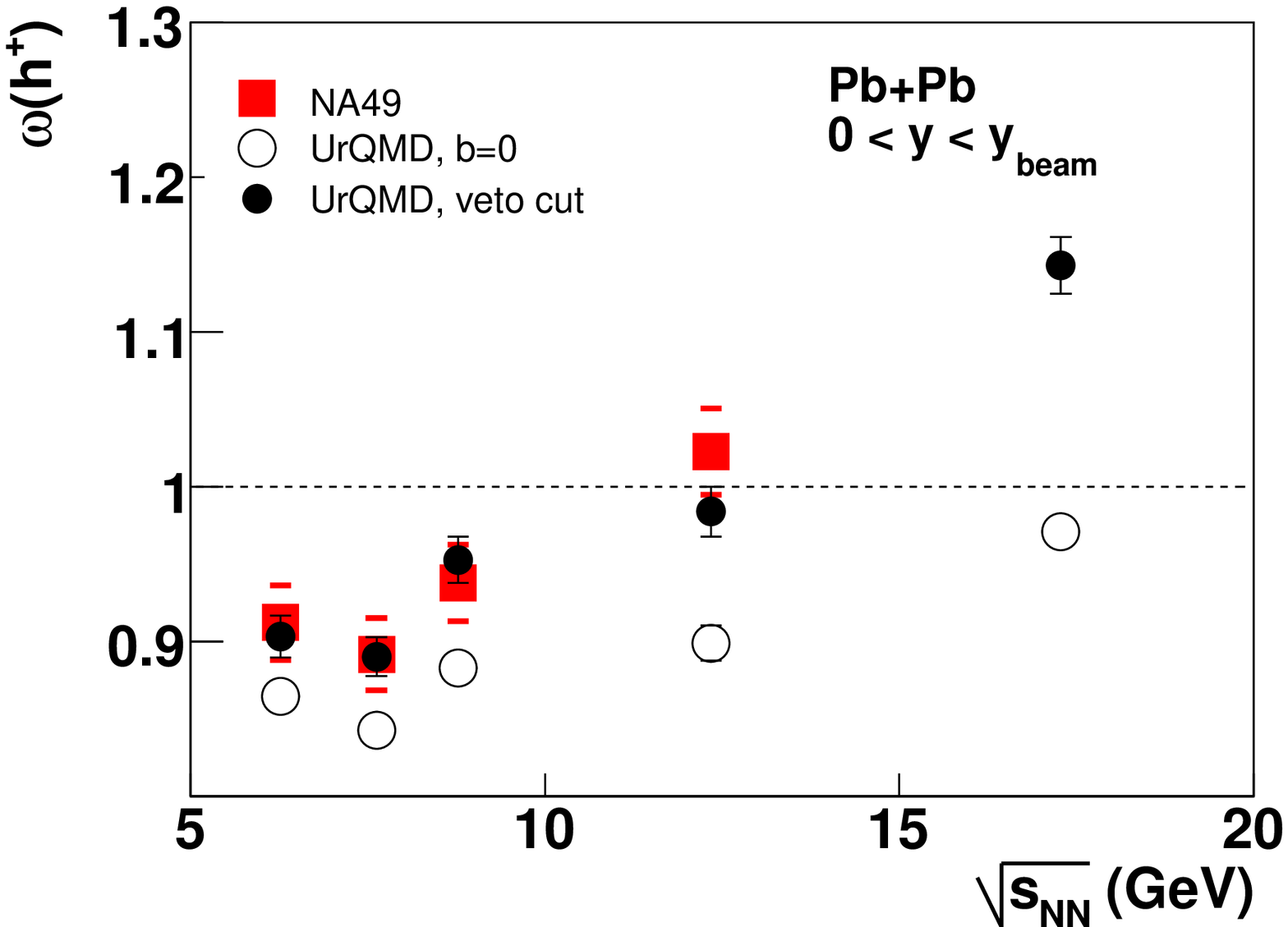}
\includegraphics[height=3.5cm]{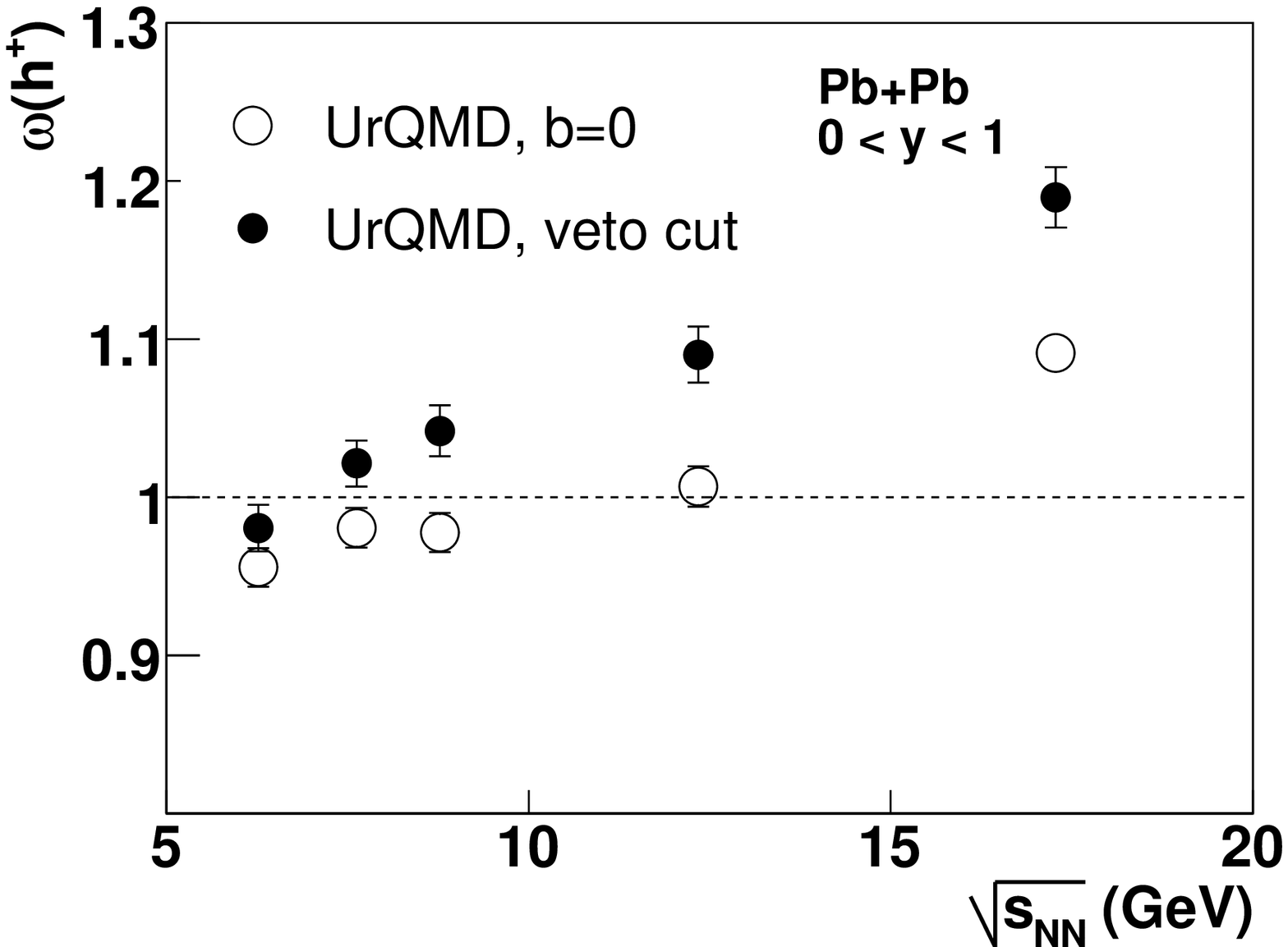}
\includegraphics[height=3.5cm]{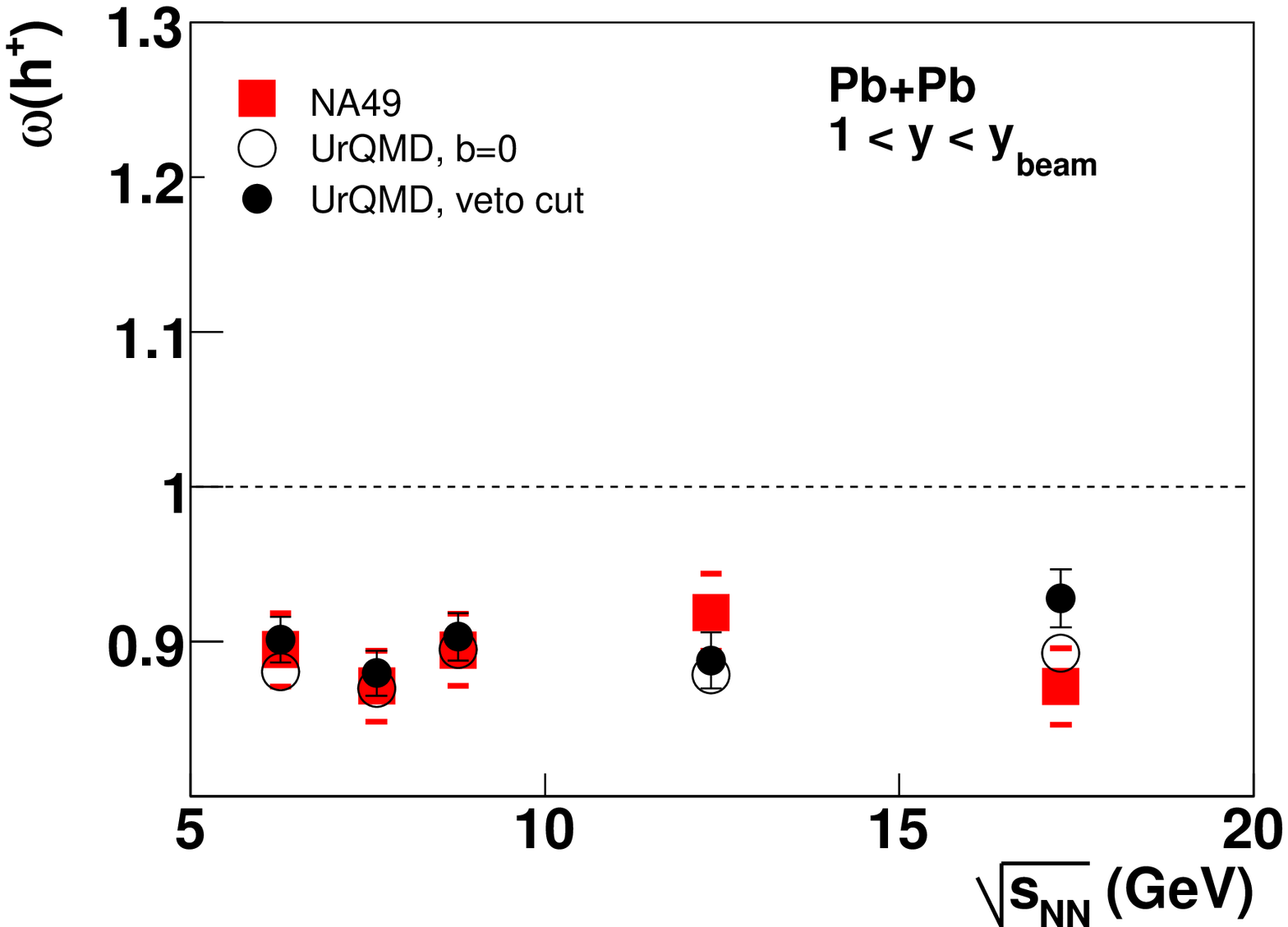}
\caption{\label{w_hp}Scaled variance of positively charged hadrons produced in central Pb+Pb collisions as a function of collision energy. 
Left: full experimental acceptance, middle: midrapidity, right: forward rapidity.}
\end{figure}

\begin{figure}
\includegraphics[height=3.5cm]{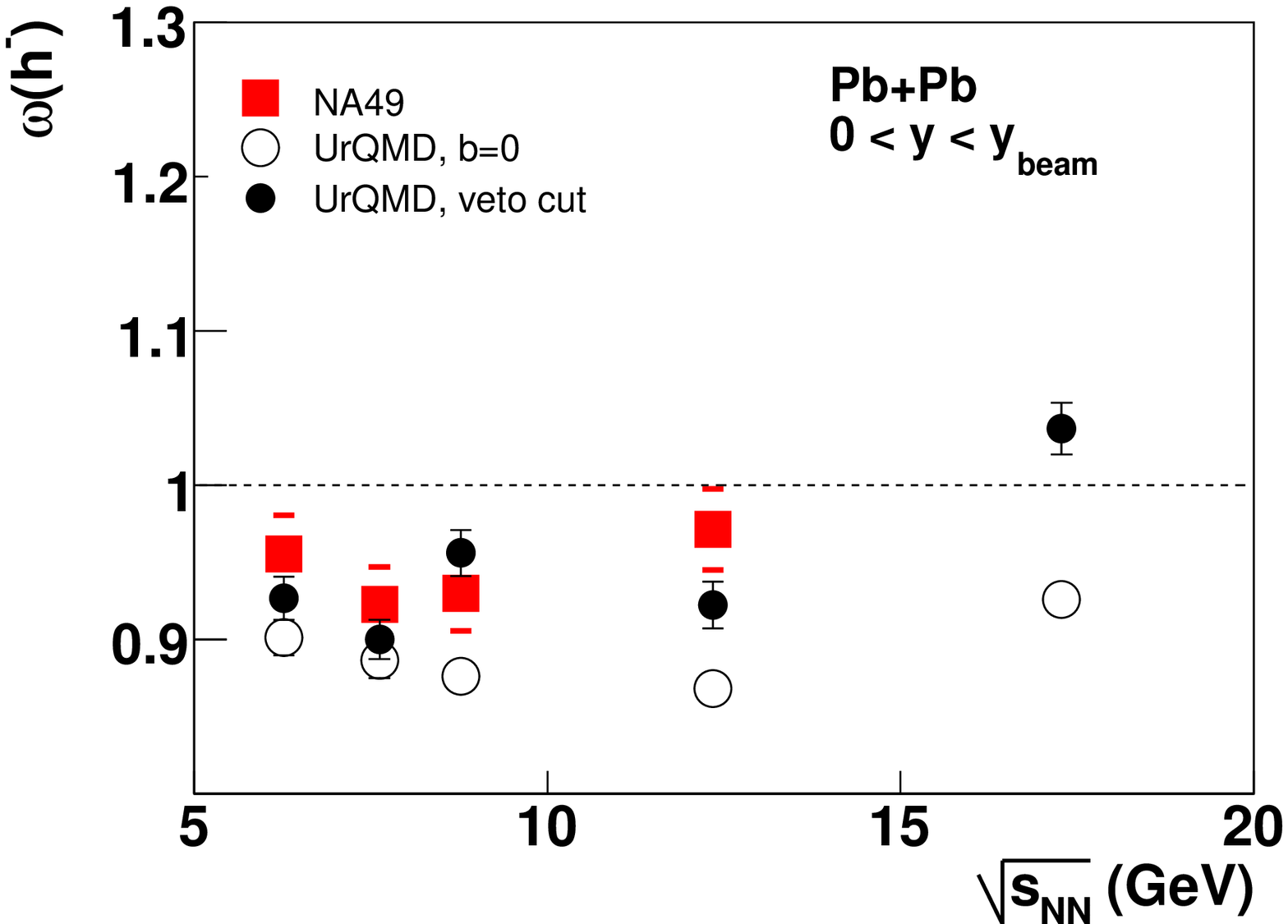}
\includegraphics[height=3.5cm]{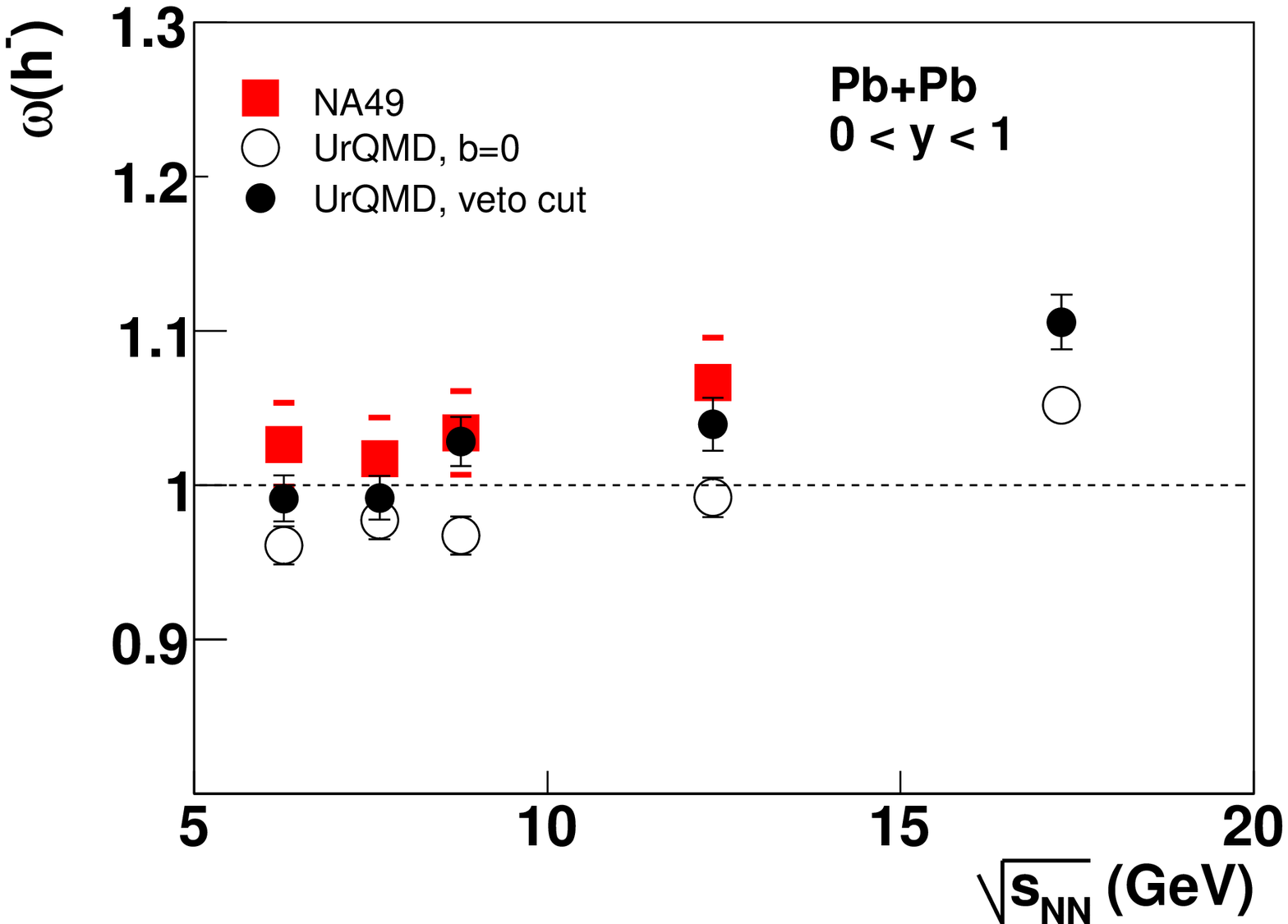}
\includegraphics[height=3.5cm]{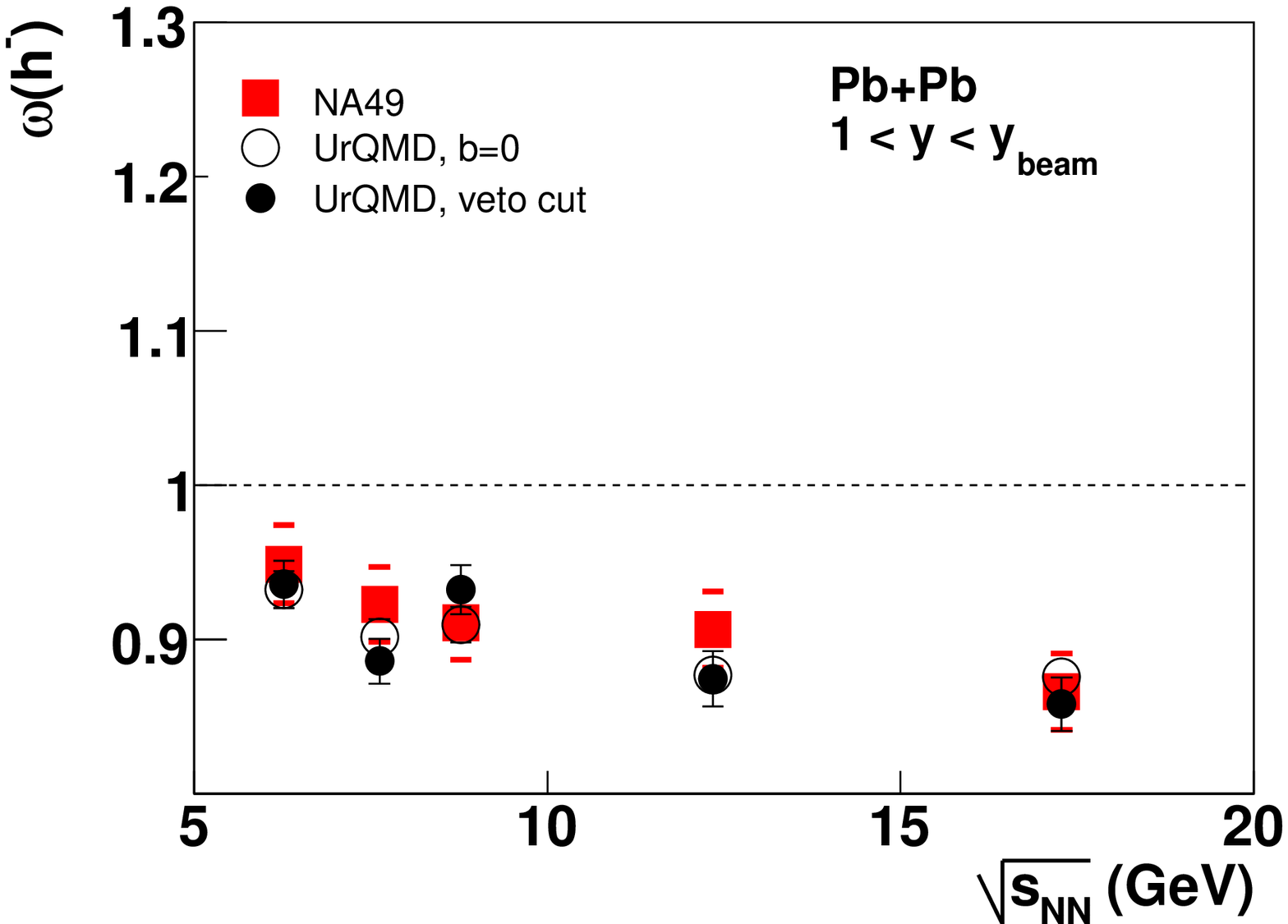}
\caption{\label{w_hm}Scaled variance of negatively charged hadrons produced in central Pb+Pb collisions as a function of collision energy. 
Left: full experimental acceptance, middle: midrapidity, right: forward rapidity.}
\end{figure}

\begin{figure}
\includegraphics[height=3.5cm]{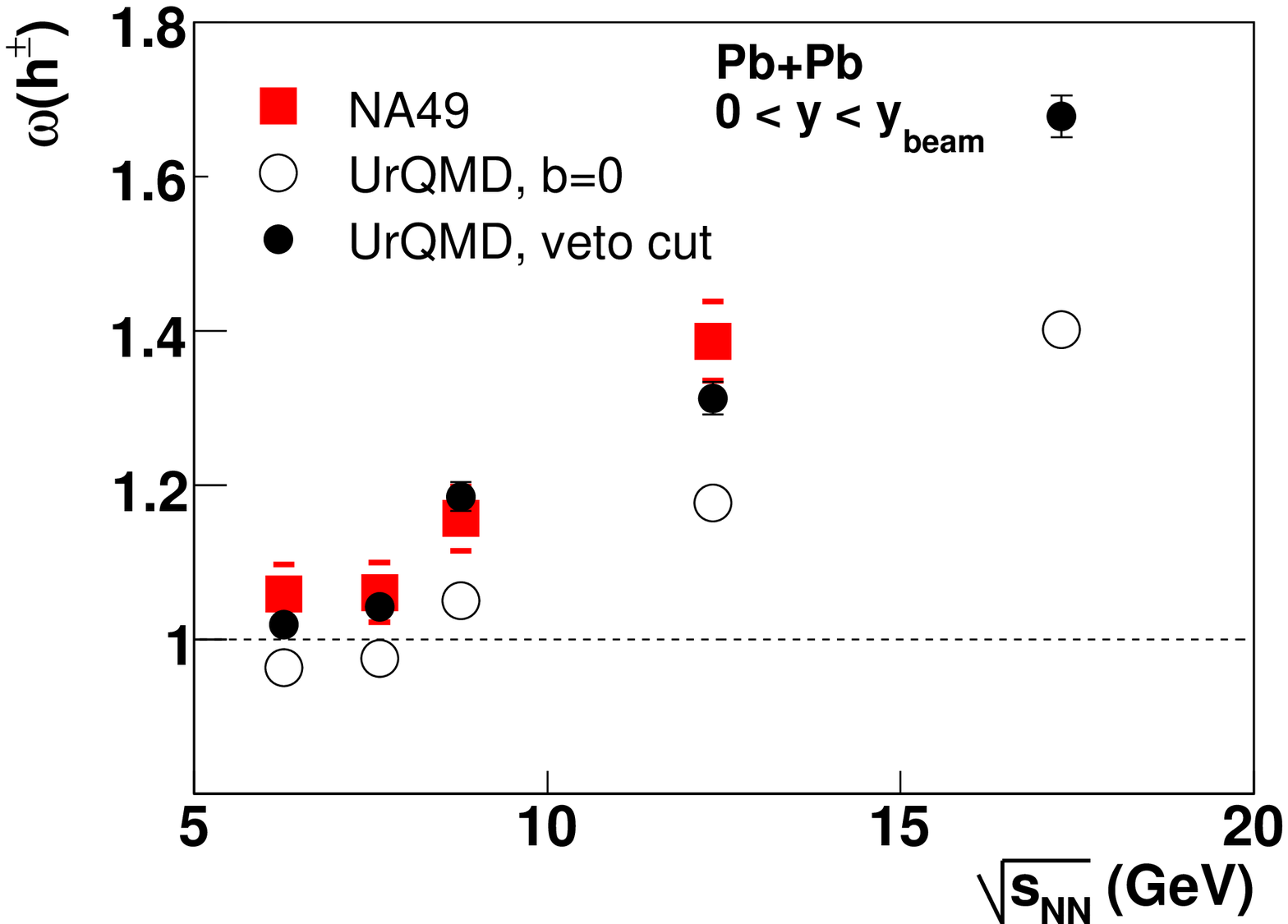}
\includegraphics[height=3.5cm]{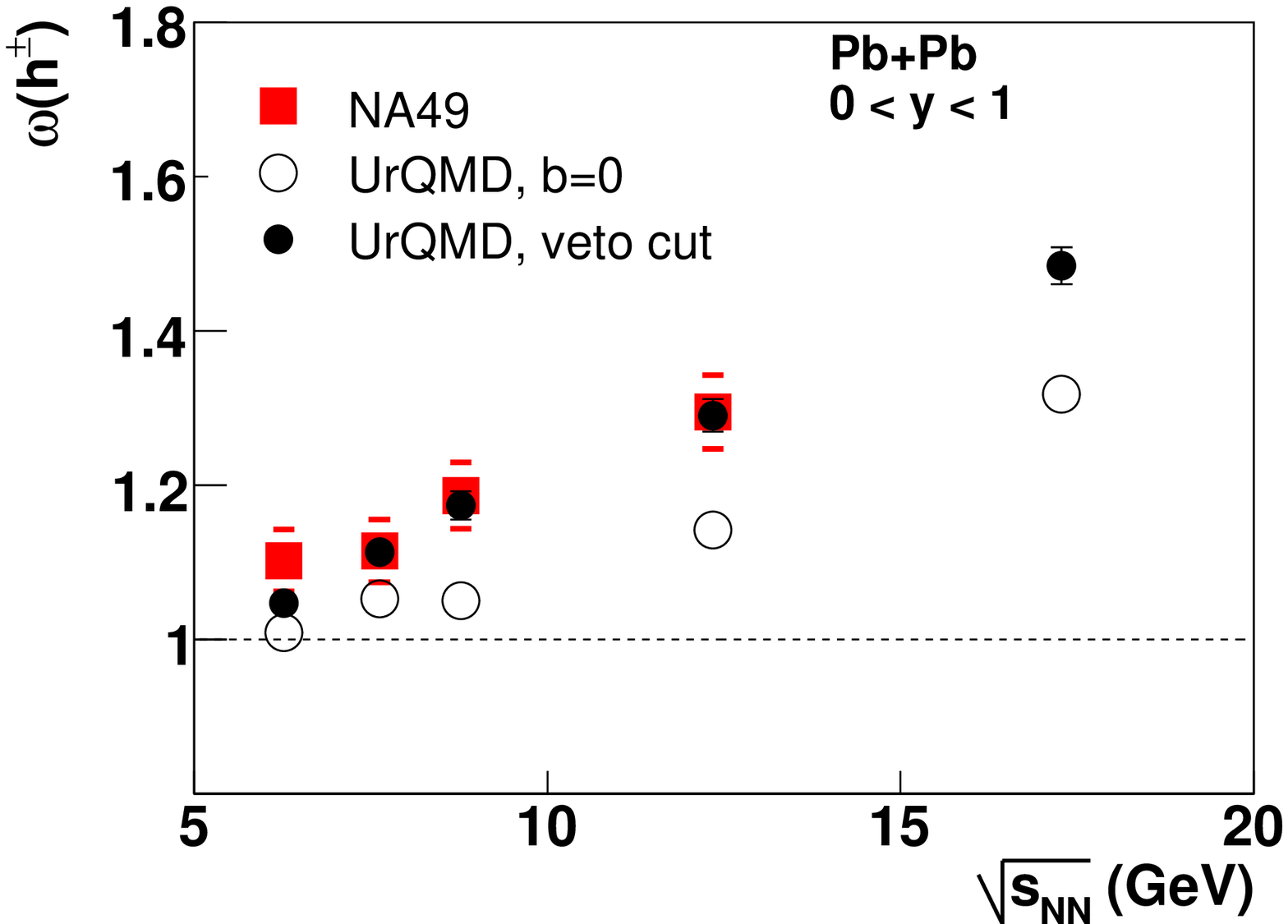}
\includegraphics[height=3.5cm]{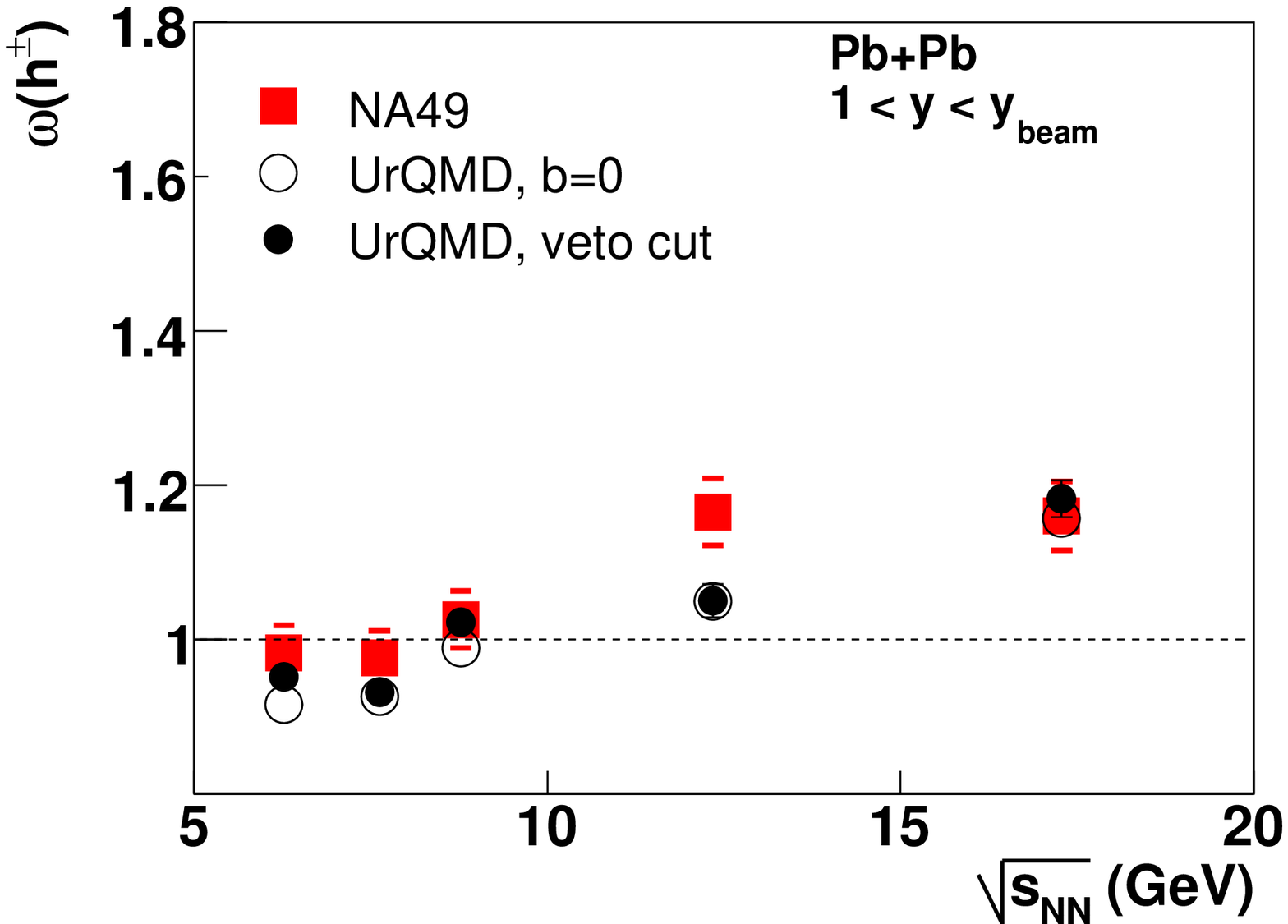}
\caption{\label{w_hpm}Scaled variance of all charged hadrons produced in central Pb+Pb collisions as a function of collision energy. 
Left: full experimental acceptance, middle: midrapidity, right: forward rapidity.}
\end{figure}

The energy dependence of the scaled variance for negatively, positively and all charged particles for three different acceptances is
shown in figures \ref{w_hp}-\ref{w_hpm}.

For positively and negatively charged hadrons the scaled variance is similar and
smaller than 1 in the forward acceptance at all energies. At midrapidity, it is larger than 1. 
For all charged particles the scaled variance is higher than for one single charge.

No significant structure or non-monotonous behaviour at a specific energy is observed.

%\subsection{Rapidity Dependence}

\begin{figure}
\includegraphics[height=3.5cm]{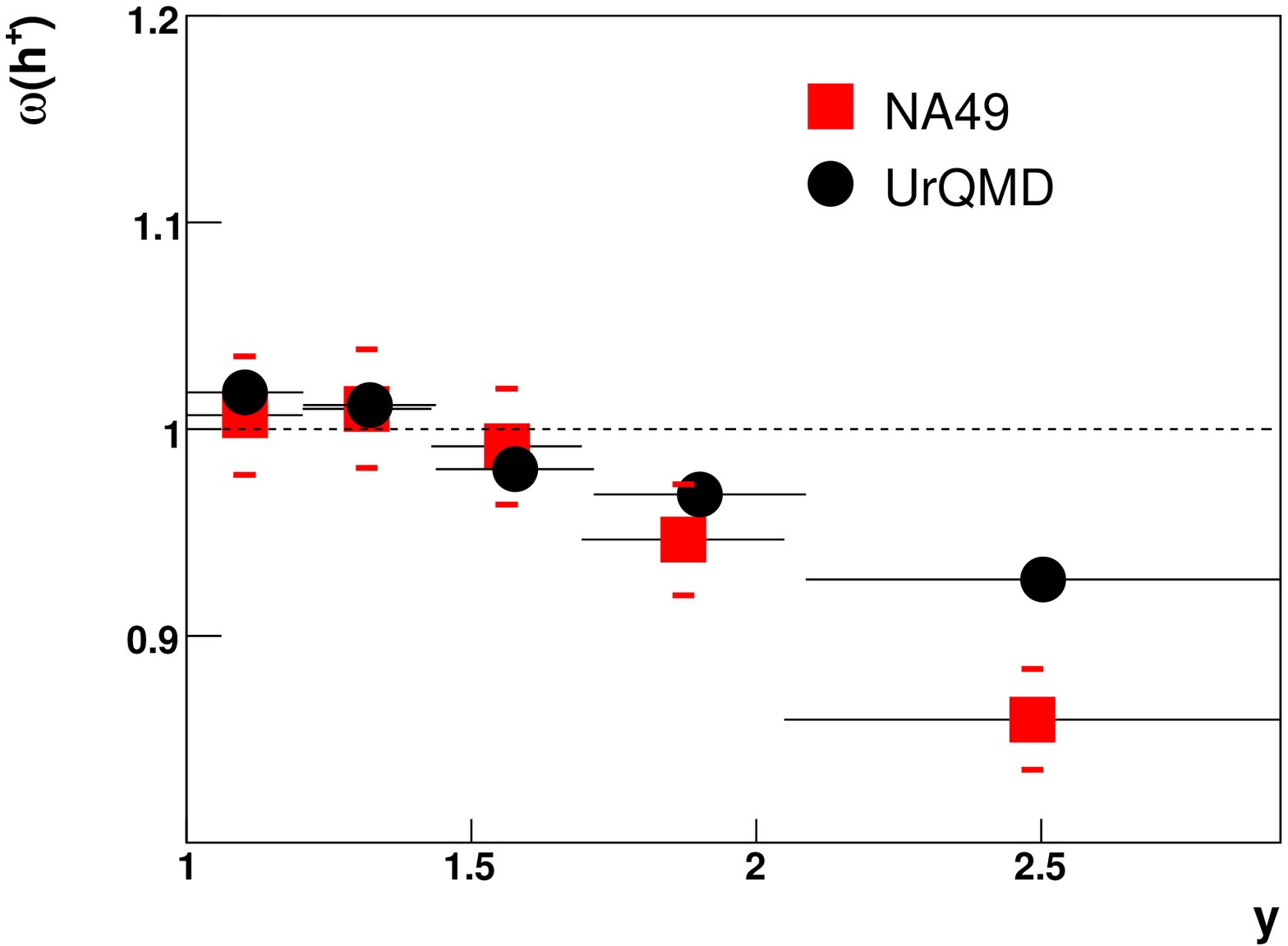}
\includegraphics[height=3.5cm]{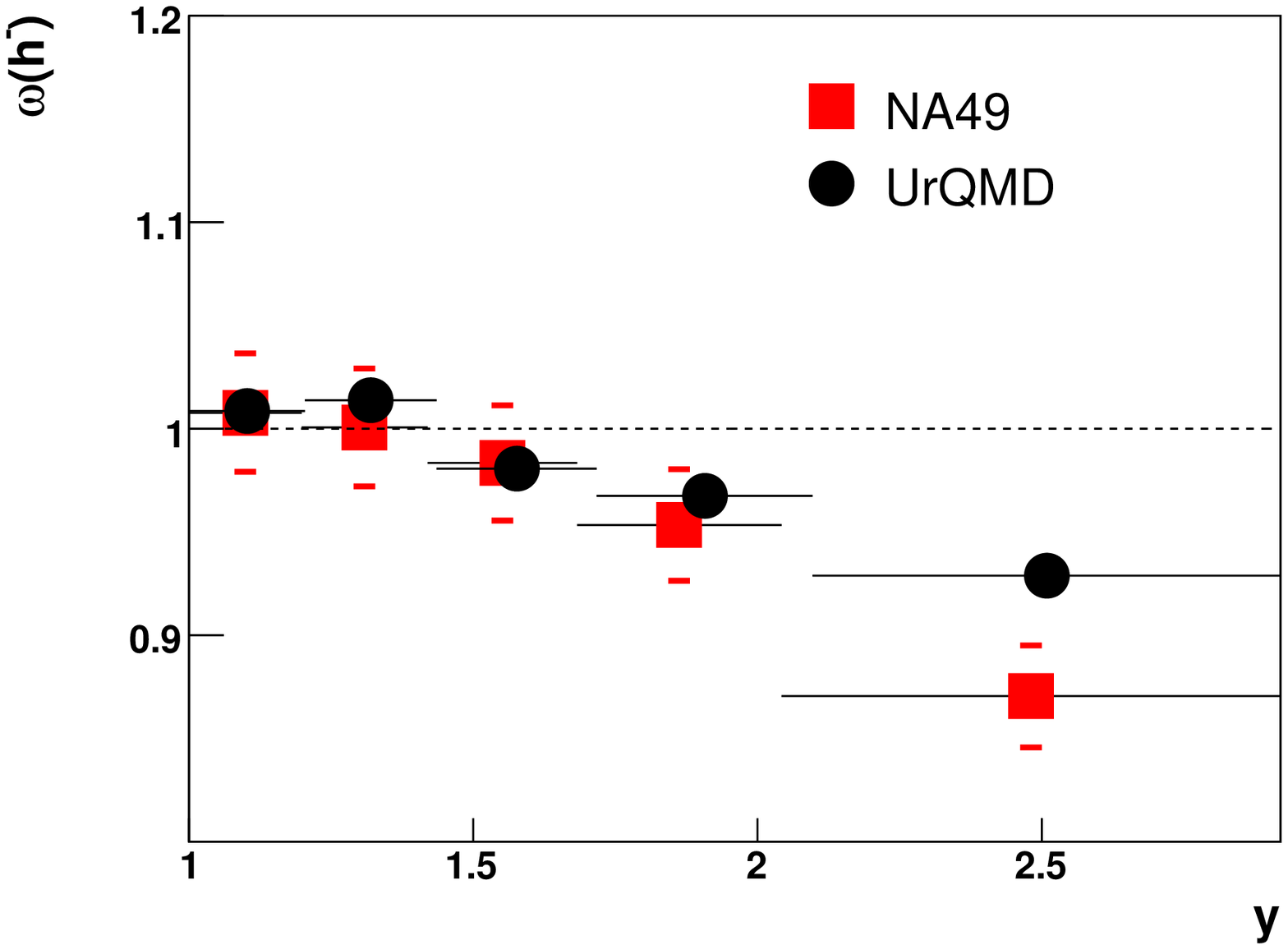}
\includegraphics[height=3.5cm]{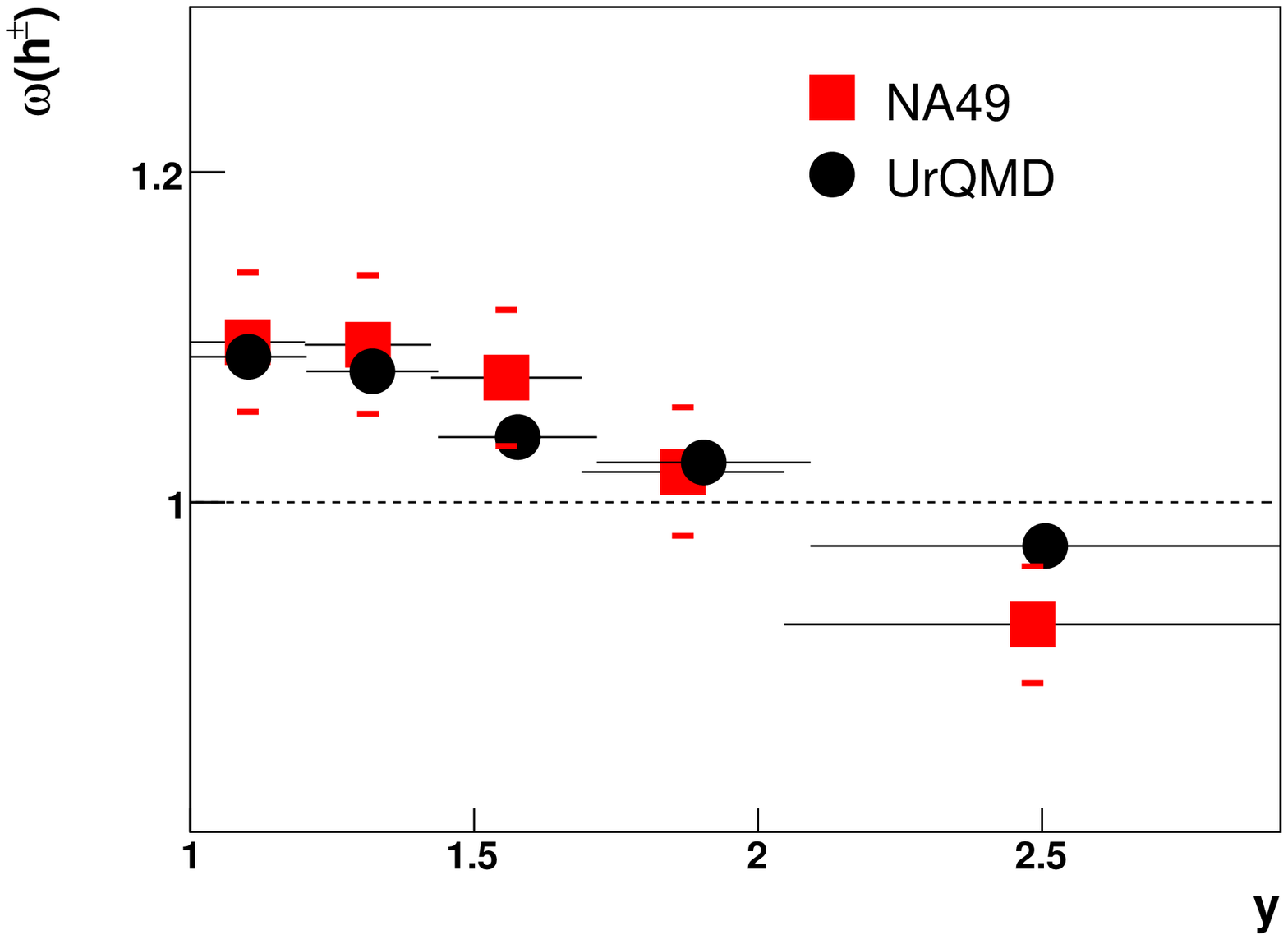}
\caption{\label{ydep}Rapidity dependence of the scaled variance of positively (left), negatively (middle) and all charged hadrons 
(right) in central Pb+Pb collisions 
at $158A$ GeV. The rapidity bins are constructed in such 
a way that the mean multiplicity in each bin is the same.}
\end{figure}

The rapidity dependence of the scaled variance 
for $158A$ GeV central Pb+Pb collisions 
is shown in figure
\ref{ydep}. 
For this energy only the rapidity interval $1<y(\pi)<y_{beam}$ is used because
the midrapidity region suffers from reconstruction inefficiency 
(see section \ref{syserr}).
In order to get rid of the "trivial" dependence of the scaled variance on 
the fraction of accepted tracks (see equation \ref{wscale})
the rapidity bins are constructed in such a way that 
the mean multiplicity in acceptance for each bin is the same. 
If there would be no correlations in momentum space, the scaled variance in figure \ref{ydep} 
would be independent of rapidity. This is not the case,
the experimental data show an increase of the scaled variance towards midrapidity for all charges. 

%\subsection{Transverse Momentum Dependence}

\begin{figure}
\includegraphics[height=3.5cm]{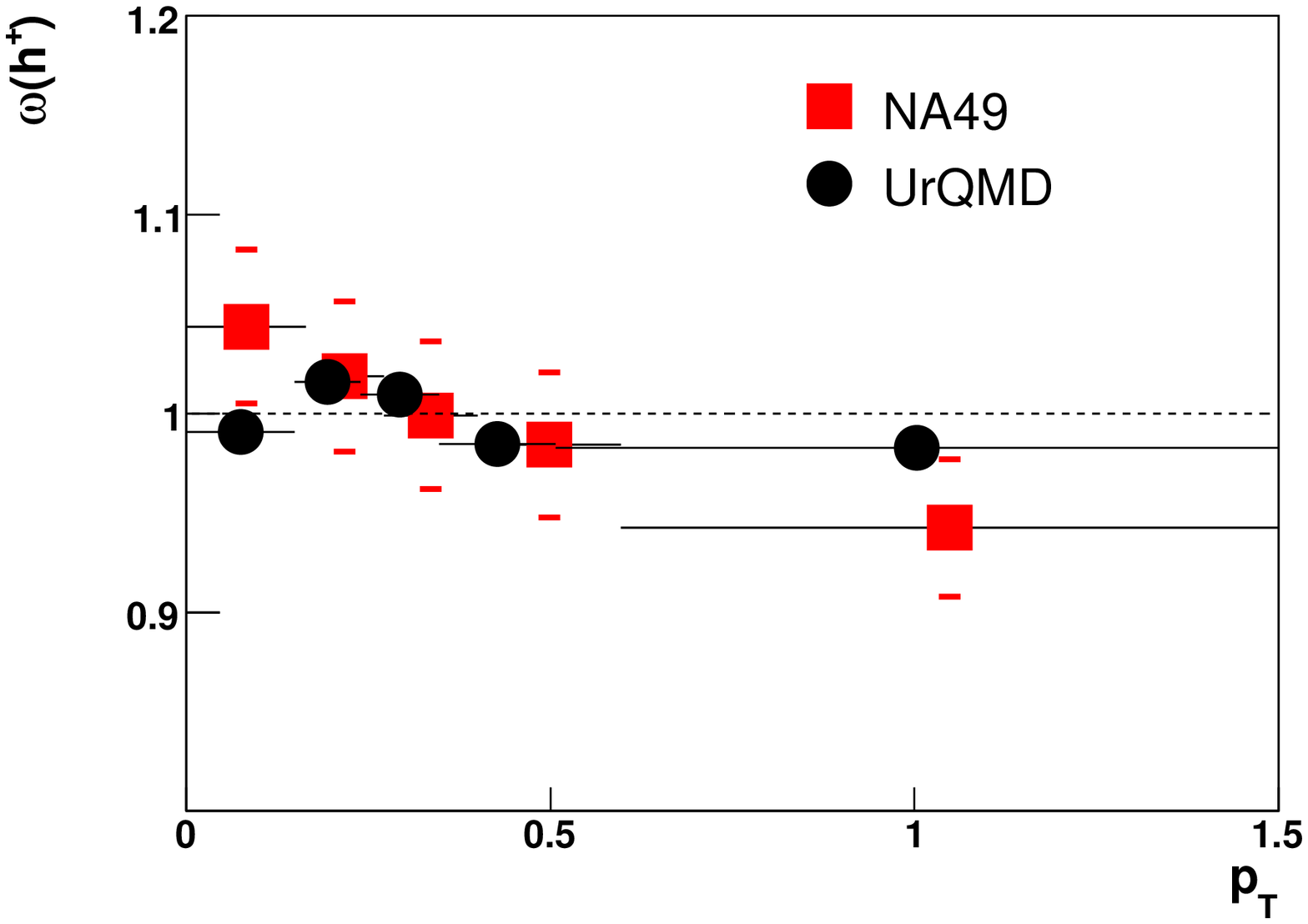}
\includegraphics[height=3.5cm]{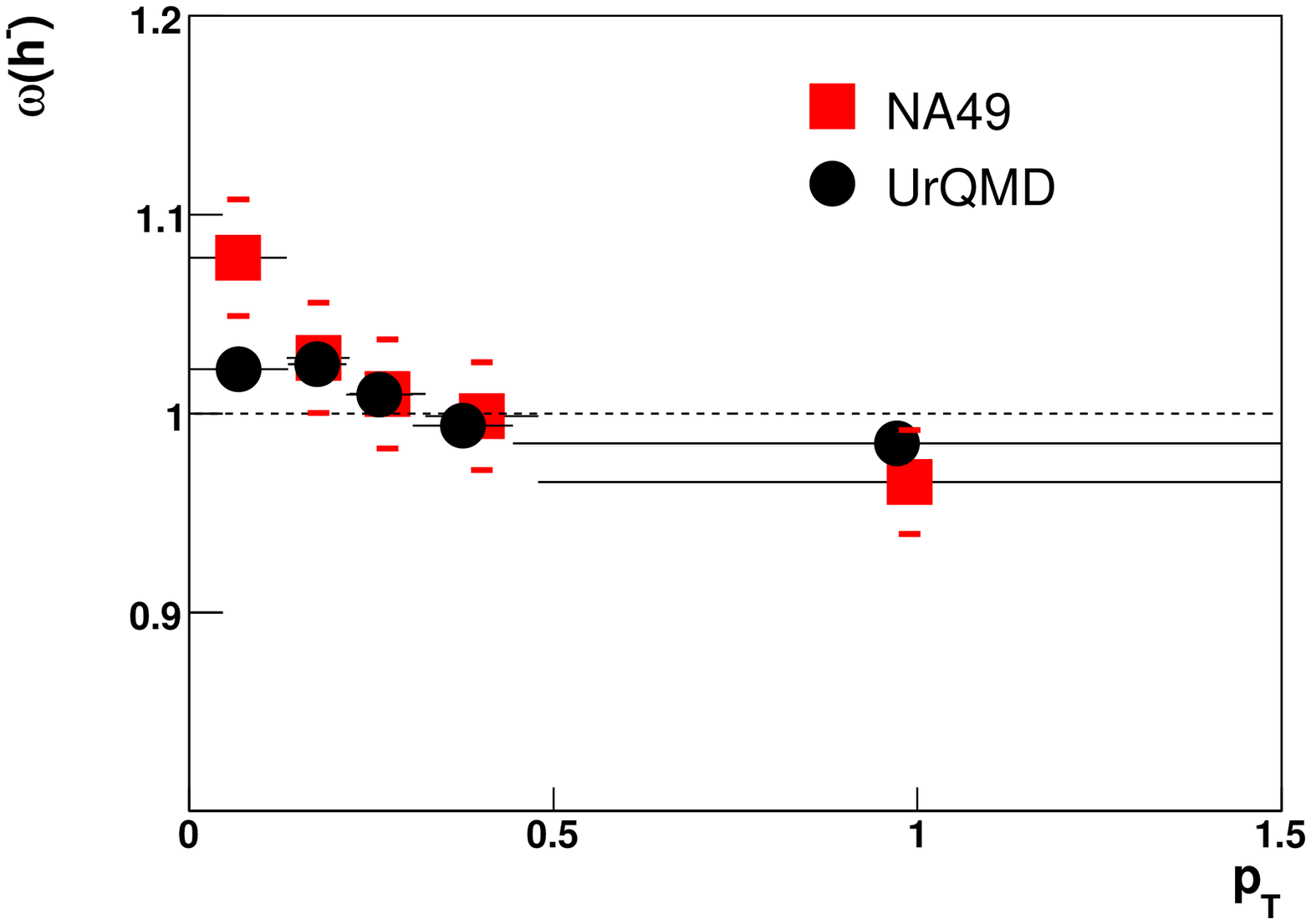}
\includegraphics[height=3.5cm]{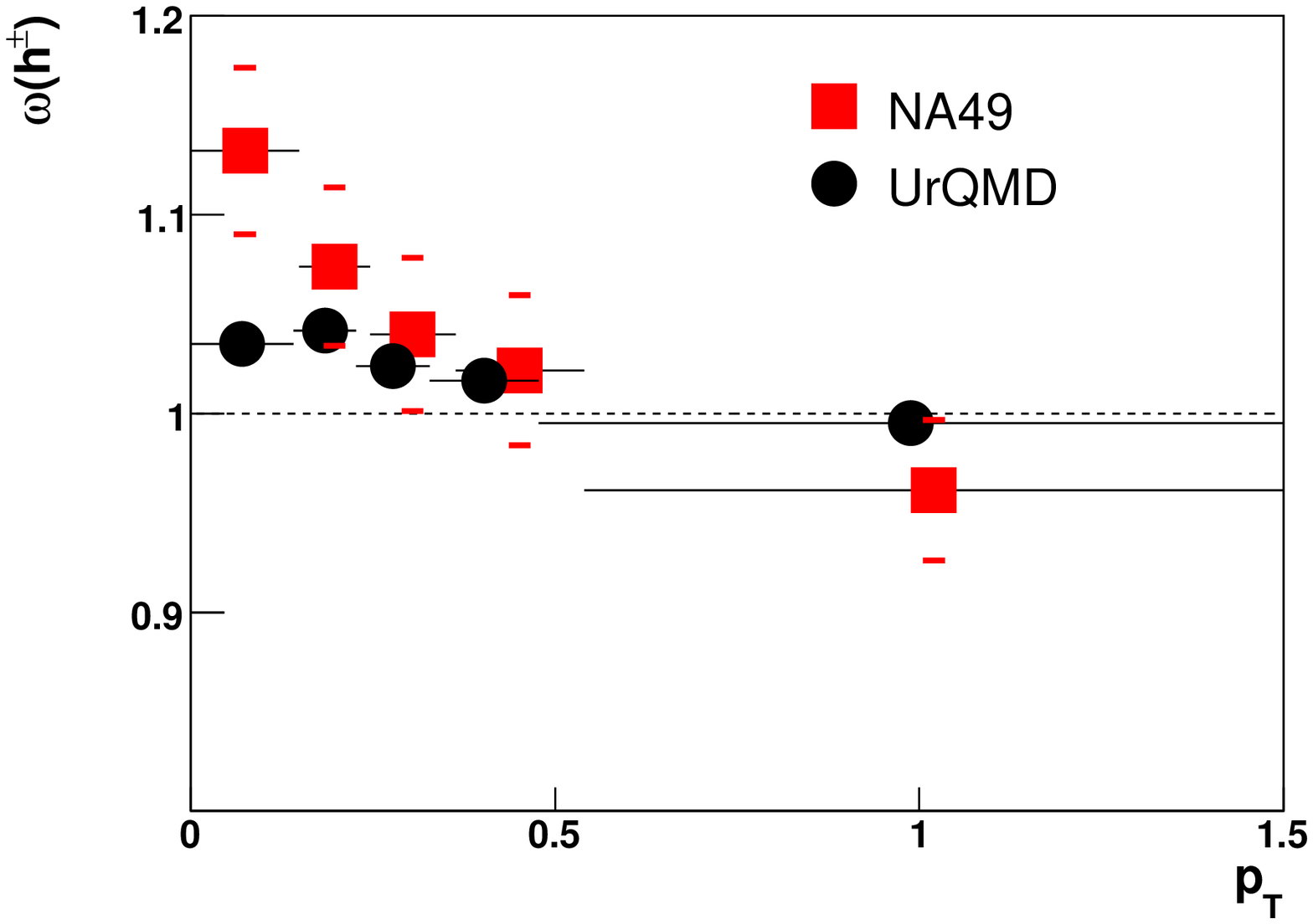}
\caption{\label{ptdep}Transverse momentum dependence of the scaled variance of positive (left), negative (middle) and all charged hadrons (right)
in the rapidity interval $1.25<y(\pi)<1.75$ in central Pb+Pb collisions at $158A$ GeV. 
}
\end{figure}

The transverse momentum dependence of the scaled variance at top SPS energy is shown in figure \ref{ptdep}. 
Only a small rapidity interval in the forward acceptance
($1.25<y(\pi)<1.75$) is used for this study. A larger rapidity interval might cause a bias because the acceptance in rapidity is different for different 
transverse momenta.

The scaled variance increases with decreasing transverse momentum. 

\section{Model Comparison}\label{c_modcomp}

\subsection{Hadron-Resonance Gas Model}\label{c_statmod}

\begin{figure}
\includegraphics[height=3.3cm]{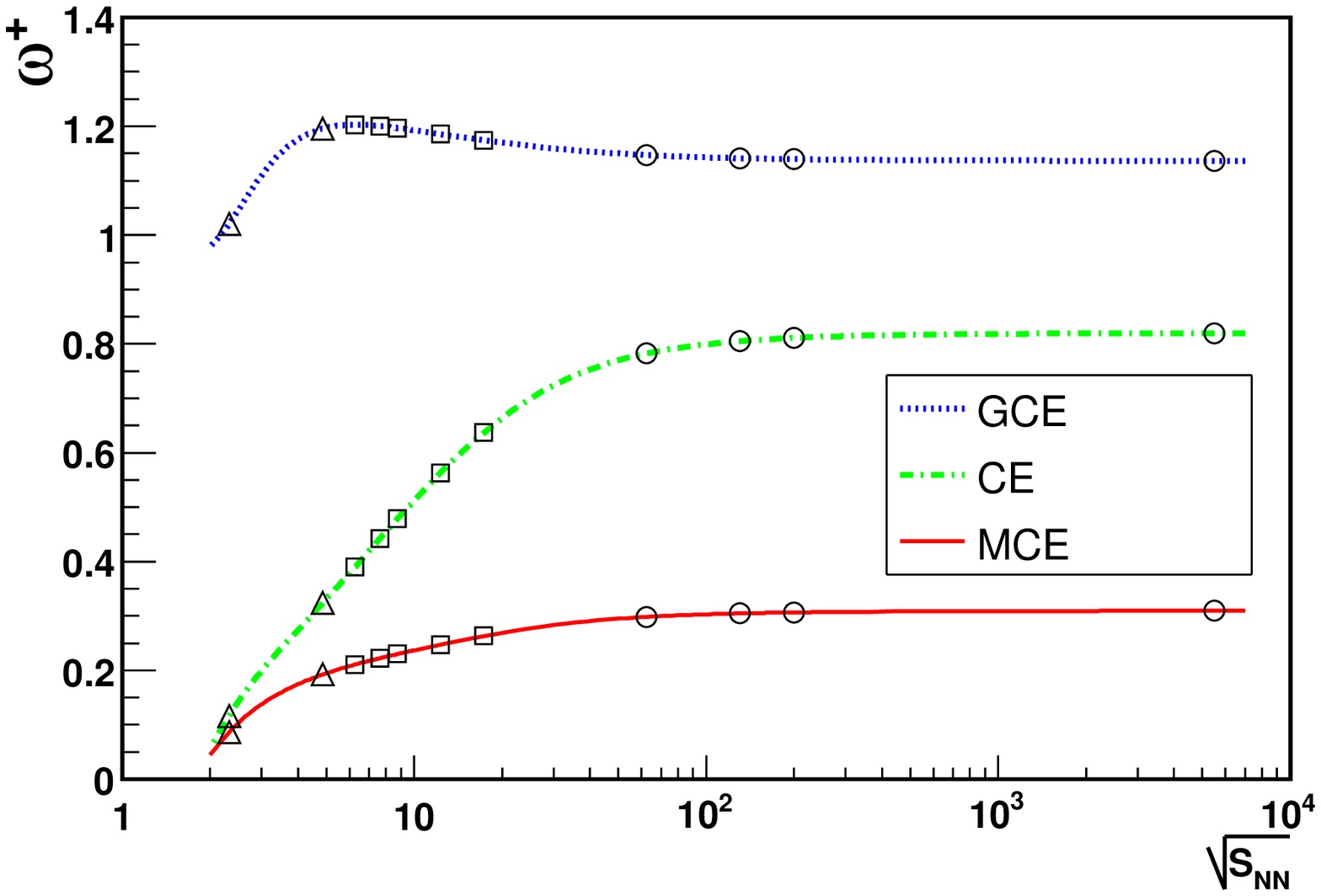}
\includegraphics[height=3.3cm]{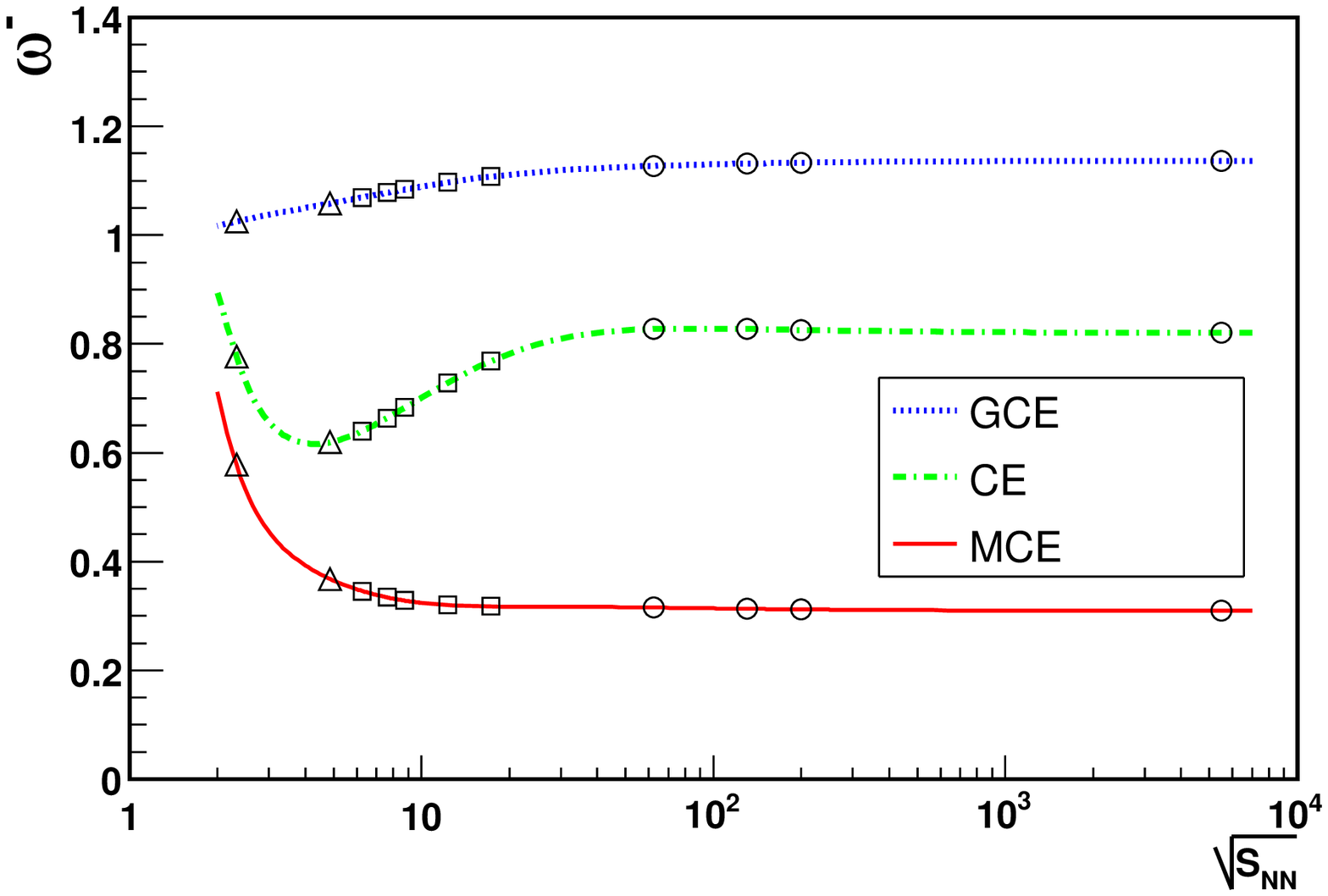}
\includegraphics[height=3.3cm]{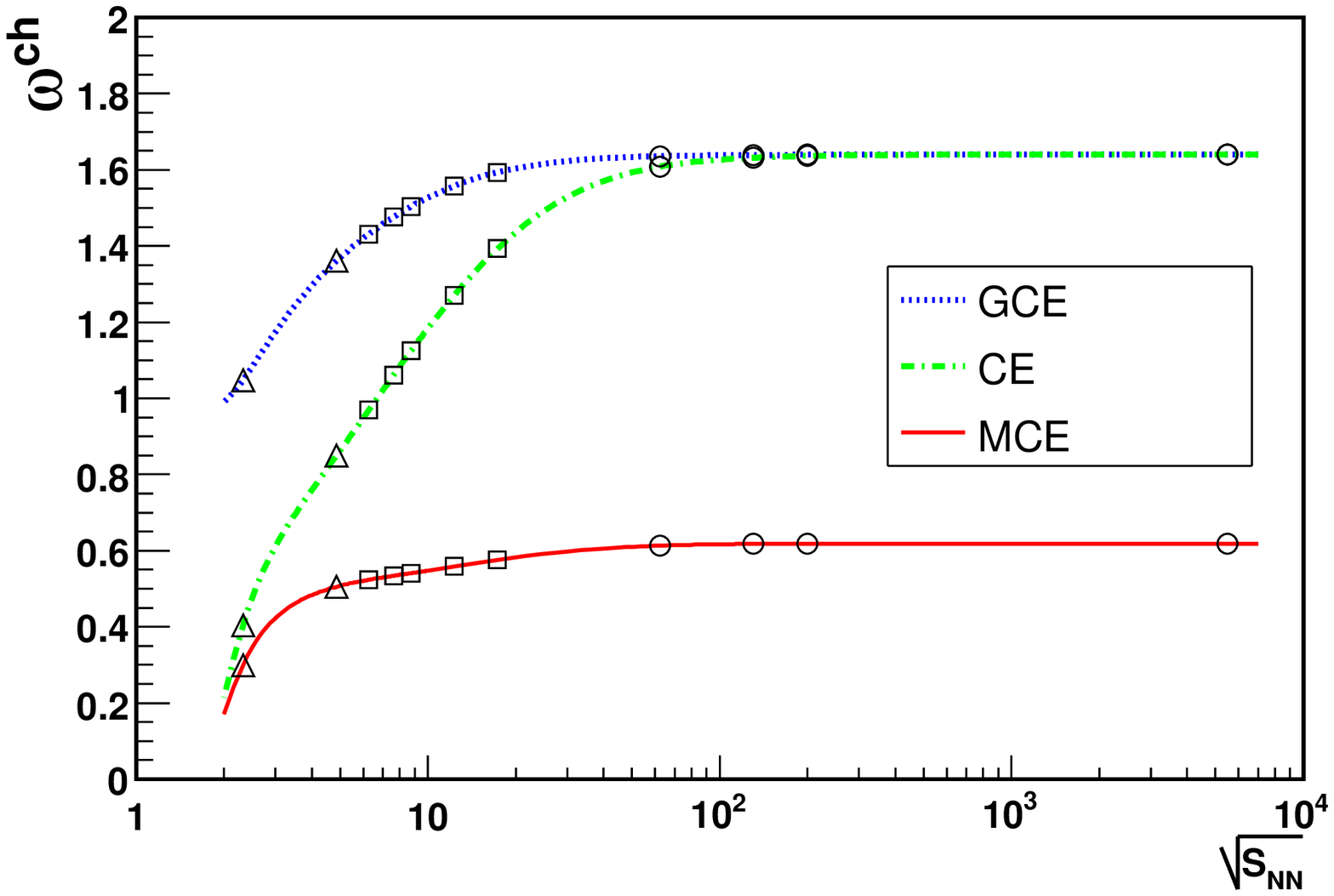}
\caption{\label{w_statmod_4pi}Predictions of a hadron-resonance 
gas model~\cite{Begun:2006uu} for the scaled variance 
in full phase space of different statistical ensembles for positively (left),
negatively and all charged hadrons (right) produced in central Pb+Pb collisions as a function of collision energy.}
\end{figure}

In a hadron gas model statistical equilibrium is assumed.
In \cite{Begun:2006uu} the fluctuations of particle multiplicity in full phase-space 
were calculated using 
the three different statistical ensembles. 
Quantum-statistical-effects and resonance decays are included in this model.
The scaled variance of negatively charged hadrons is shown in figure~\ref{w_statmod_4pi}.

The scaled variance is largest in the grand-canonical ensemble, 
where all conservation laws are fulfilled only on average, not on an event-by-event basis.
In the canonical ensemble, the electric charge, 
the baryon number and the strangeness is conserved in each event. 
These constraints strongly suppress the
multiplicity fluctuations. 
In the micro-canonical ensemble also energy and momentum are conserved in each collision. 
The fluctuations are the smallest 
in this ensemble.

\begin{figure}
\begin{center}
\includegraphics[height=4cm]{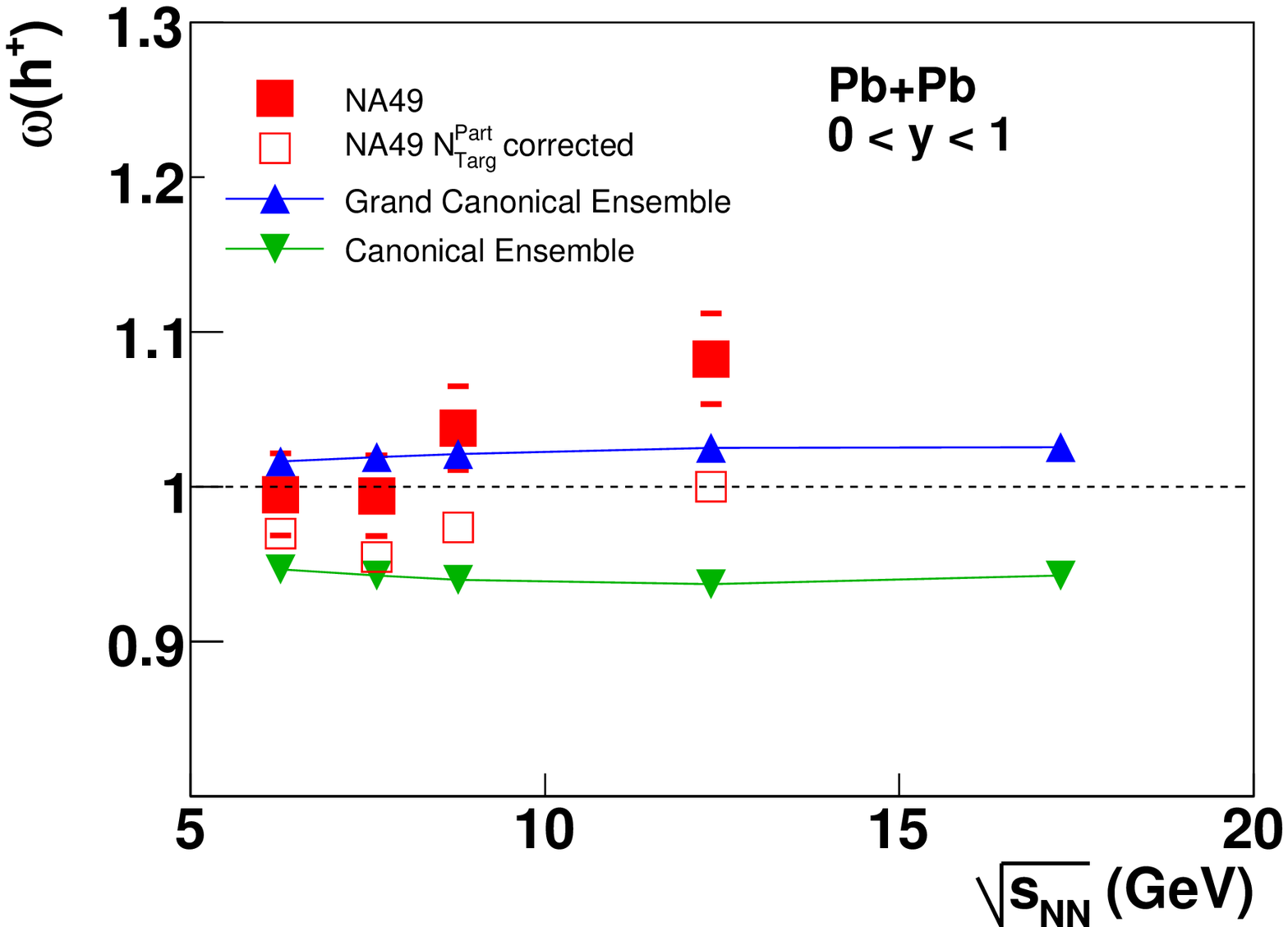}
\includegraphics[height=4cm]{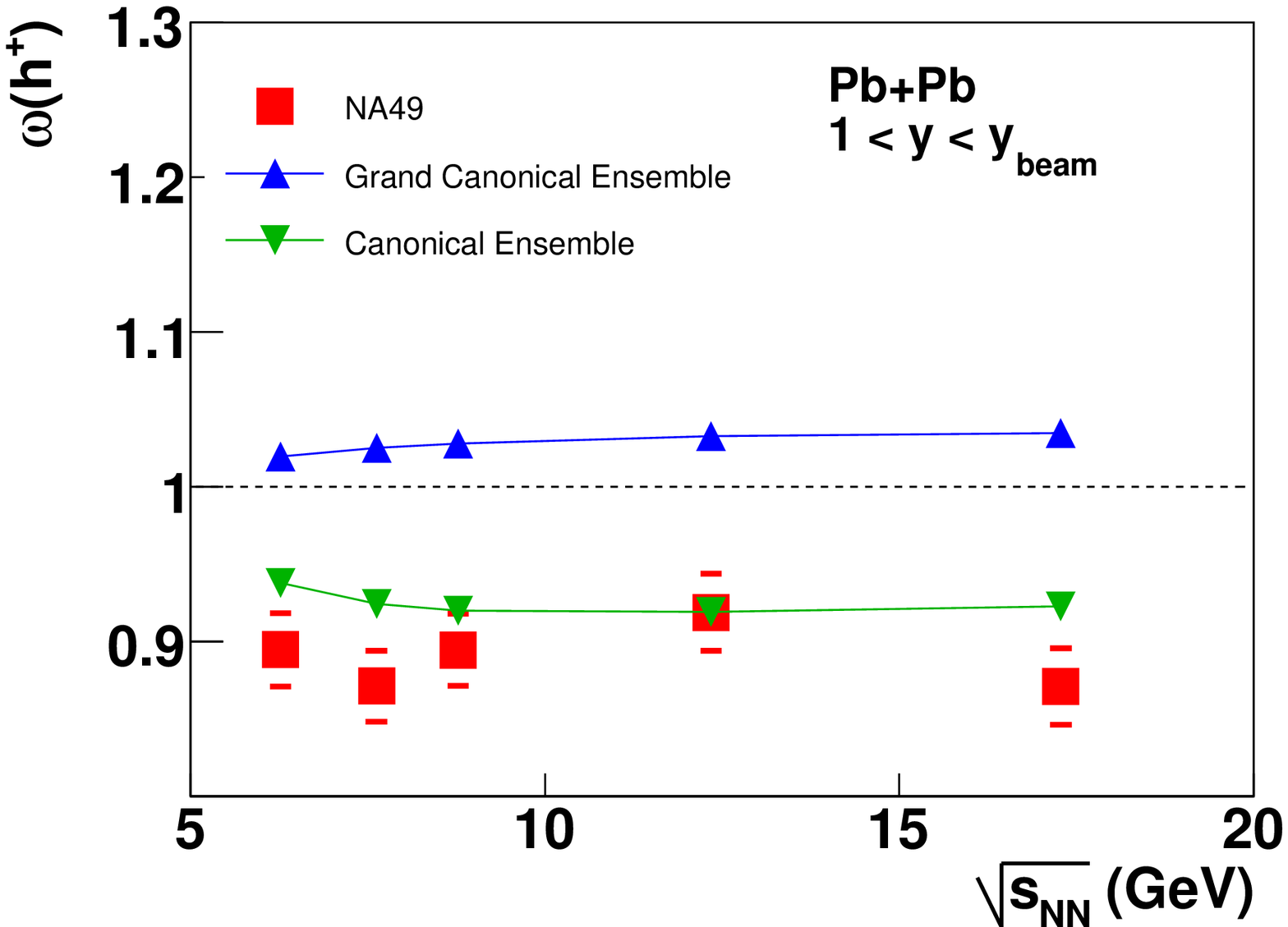}\\
\includegraphics[height=4cm]{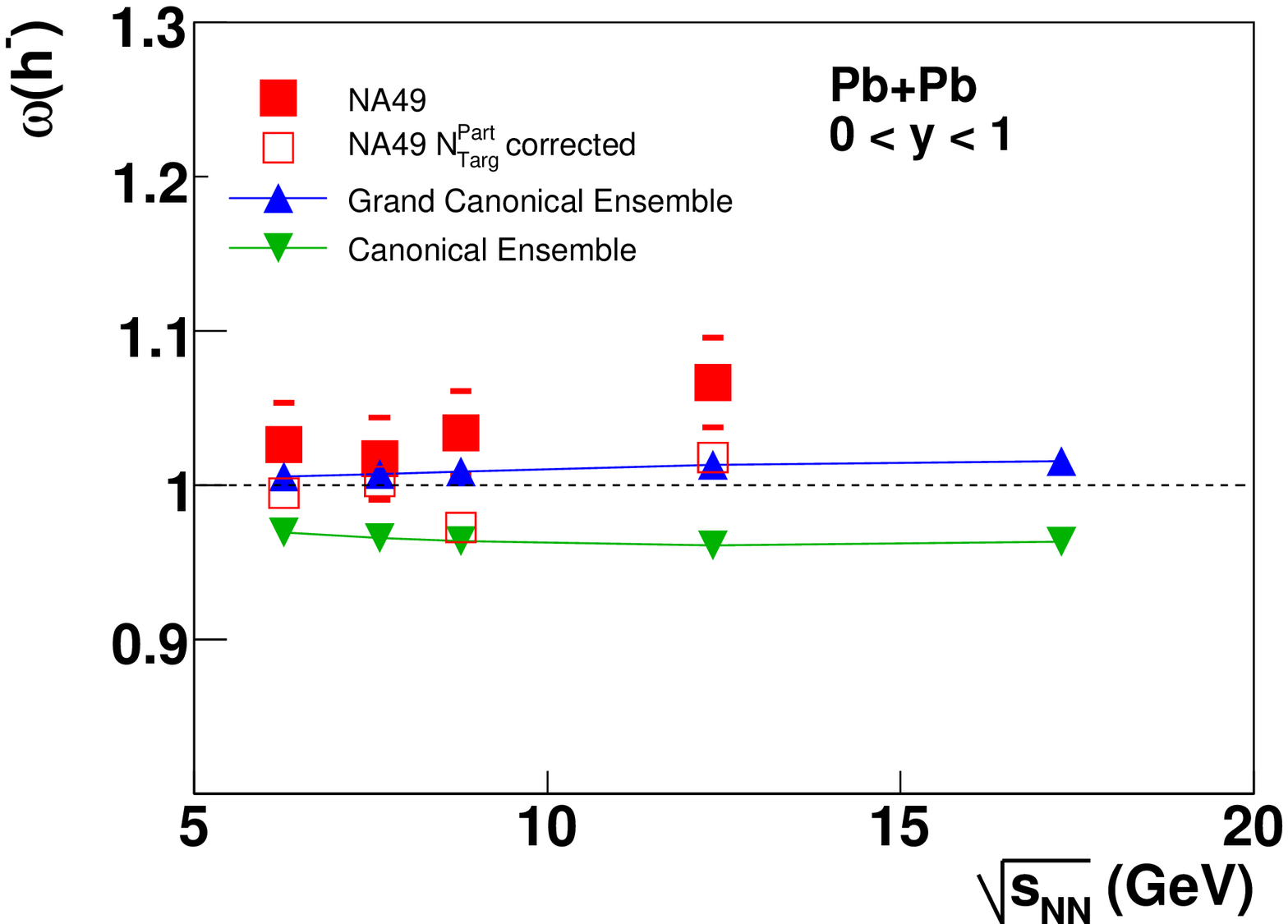}
\includegraphics[height=4cm]{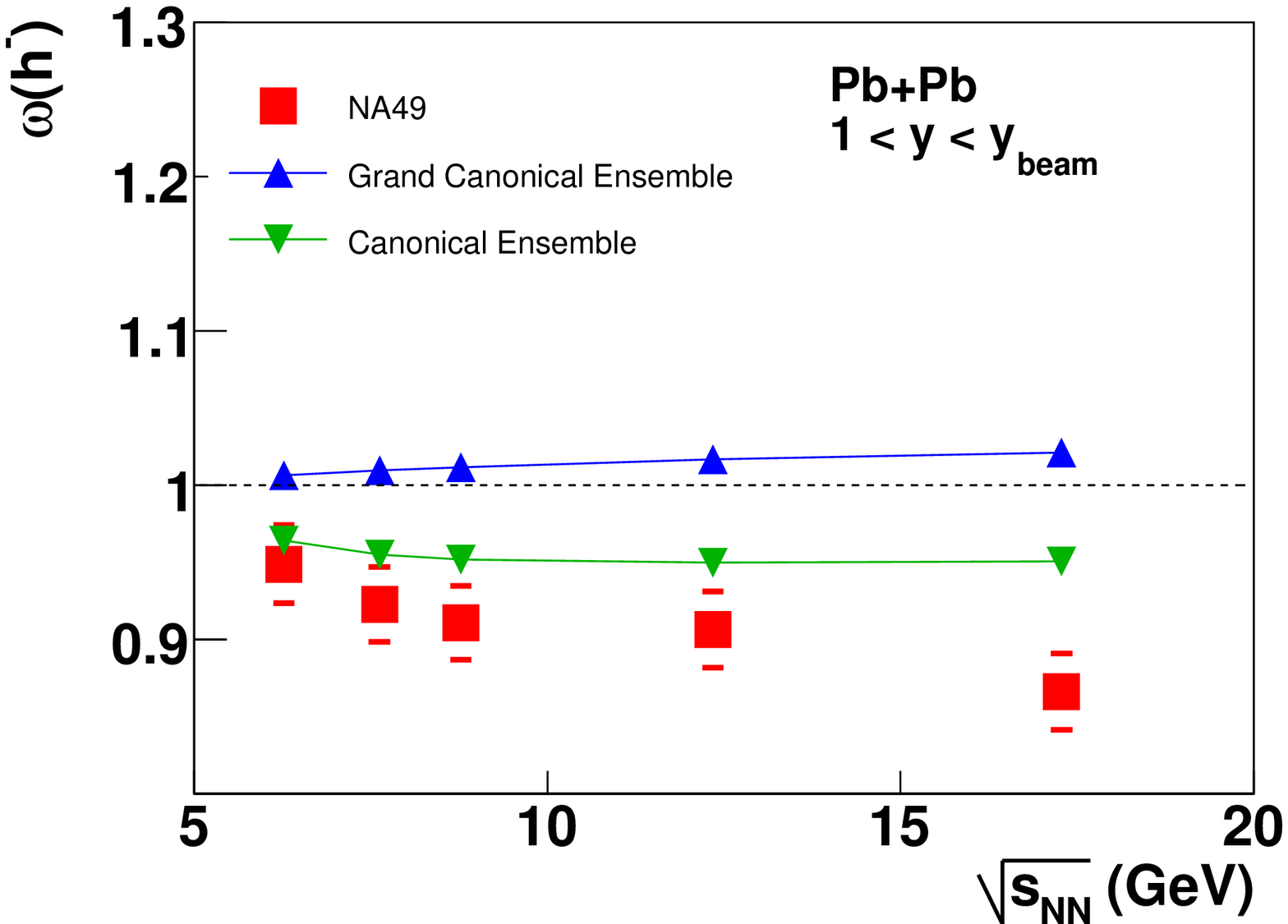}\\
\end{center}
\caption{\label{w_statmod}Scaled variance of positively (top) and negatively charged (bottom) hadrons produced in central 
Pb+Pb collisions as a function of collision 
energy
in midrapidity (left) and forward (right) 
acceptance in comparison to predictions of 
a grand canonical and canonical statistical model \cite{Begun:2006uu}.}
\end{figure}

In order to compare the hadron-resonance gas model predictions with experimental data, 
the scaled variance calculated in full phase space
is extrapolated to experimental acceptance
using equation \ref{wscale} (see figure~\ref{w_statmod}).
For the micro-canonical ensemble the presence of energy and momentum 
conservation laws introduce strong correlations in momentum space, therefore 
equation \ref{wscale} is not applicable. 
Resonance decays introduce only a weak correlation in momentum space 
for positively and negatively charged hadrons, because only a small number
of resonances decay into two particles with the same charge. 
In contrast a large number of resonances decay into two oppositely charged hadrons,
therefore equation \ref{wscale} is not valid for the scaled variance of all charged hadrons.
Quantum correlations, which introduce correlations in momentum space, are expected to have a small effect on multiplicity 
fluctuations~\cite{Begun:2006uu}. 

The fluctuations of target participants are not included in the statistical model. 
UrQMD simulations (see section \ref{centsel}) suggest that this effect is 
small in the forward region, but target participant fluctuations 
would increase the scaled variance at midrapidity significantly.
The influence of target participant fluctuations is estimated by the UrQMD model and shown as
open boxes in figure \ref{w_statmod}.

At forward rapidity, the fluctuations are overpredicted by the grand canonical model. 
The canonical model is closer to data, but overpredicts them, too.
A micro-canonical ensemble predicts smaller fluctuations 
than the canonical one in $4\pi$, but a quantitative comparison with data is not possible yet, because
correlations in momentum space do not allow to extrapolate 
to experimental acceptance using equation \ref{wscale}.
At midrapidity the scaled variance in data is much higher than in the forward region. 
This is in contradiction to the grand-canonical
and canonical statistical model, because these ensembles predict 
a similar value of the scaled variance in both regions of the phase-space.
The data points at midrapidity corrected for target participant  
fluctuations lie between the predictions of the grand-canonical and canonical model.

The canonical and grand canonical statistical models predict no dependence 
of scaled variance on rapidity and $p_T$. This is in contradiction to
experimental data shown in figures \ref{ydep} and \ref{ptdep}. 
The hadron-resonance gas model in the micro-canonical ensemble 
predicts an increase of fluctuations near midrapidity and for
low $p_T$~\cite{Hauer:2007im}, as it is seen in data, as an effect of energy and momentum conservation. 

\subsection{The UrQMD Model}\label{urqmd_s}

The UrQMD (v1.3)~\cite{Bleicher:1999xi,Bass:1998ca} microscopic transport approach is based on the covariant propagation of
constituent quarks and di-quarks accompanied by mesonic and baryonic
degrees of freedom. It simulates multiple interactions of
in-going and newly produced particles, the excitation
and fragmentation of colour strings and the formation and decay of
hadronic resonances.
Towards higher energies, the treatment of sub-hadronic degrees of freedom is
of major importance.
In the present model, these degrees of freedom enter via
the introduction of a formation time for hadrons produced in the
fragmentation of strings.
A phase transition to a quark-gluon state is
not incorporated explicitly into the model dynamics. 

\begin{figure}
\includegraphics[height=3.5cm]{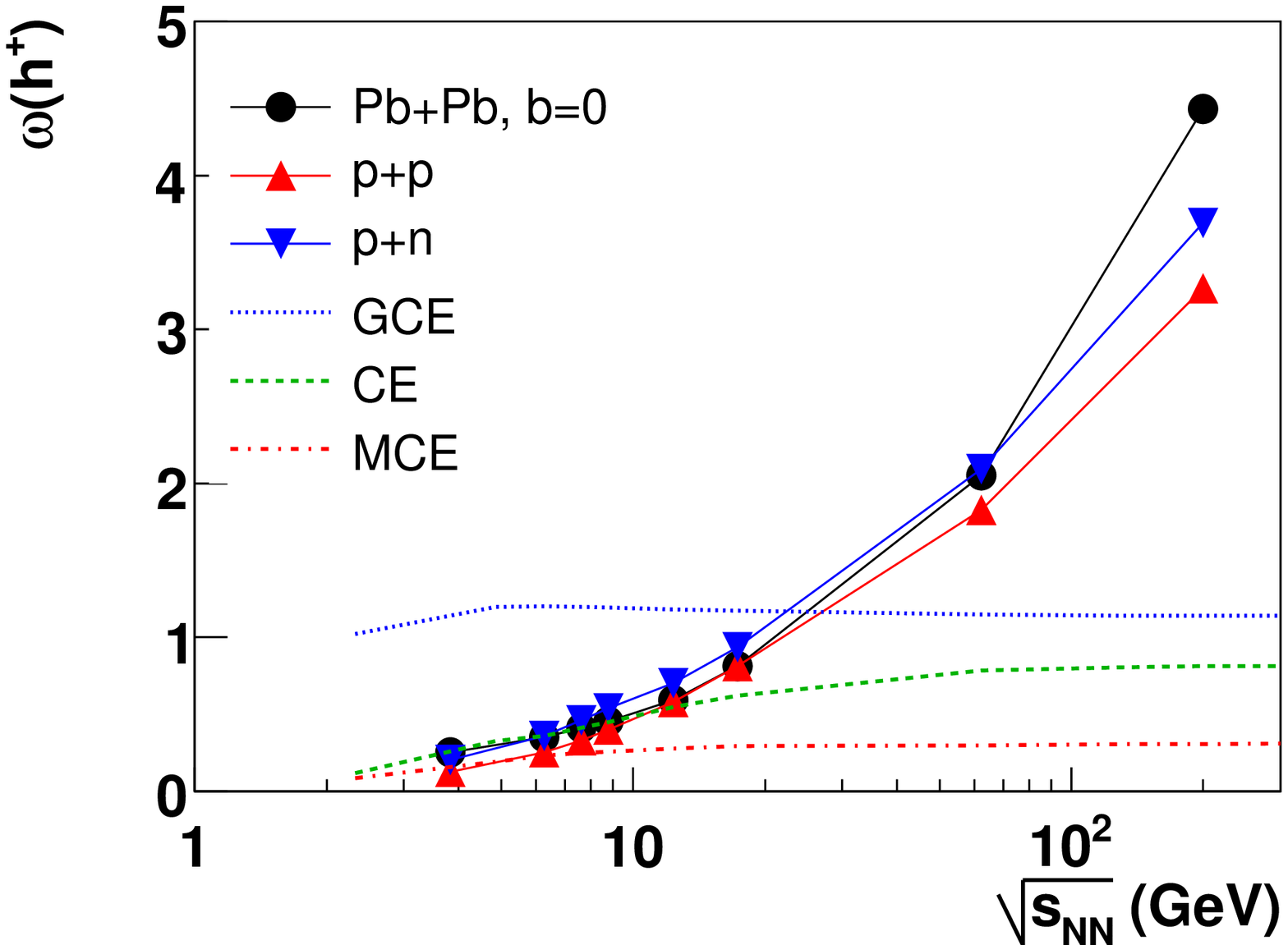}
\includegraphics[height=3.5cm]{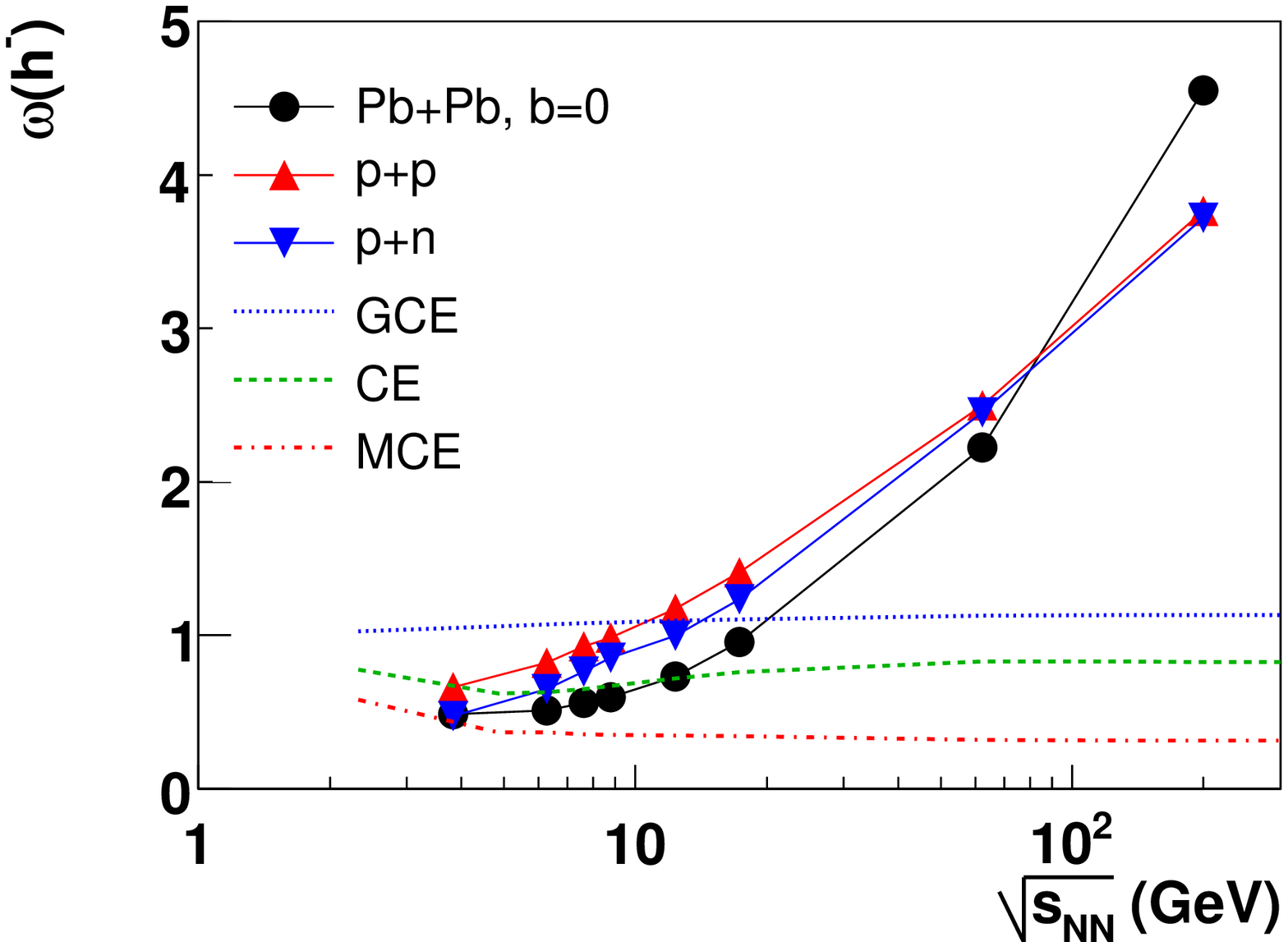}
\includegraphics[height=3.5cm]{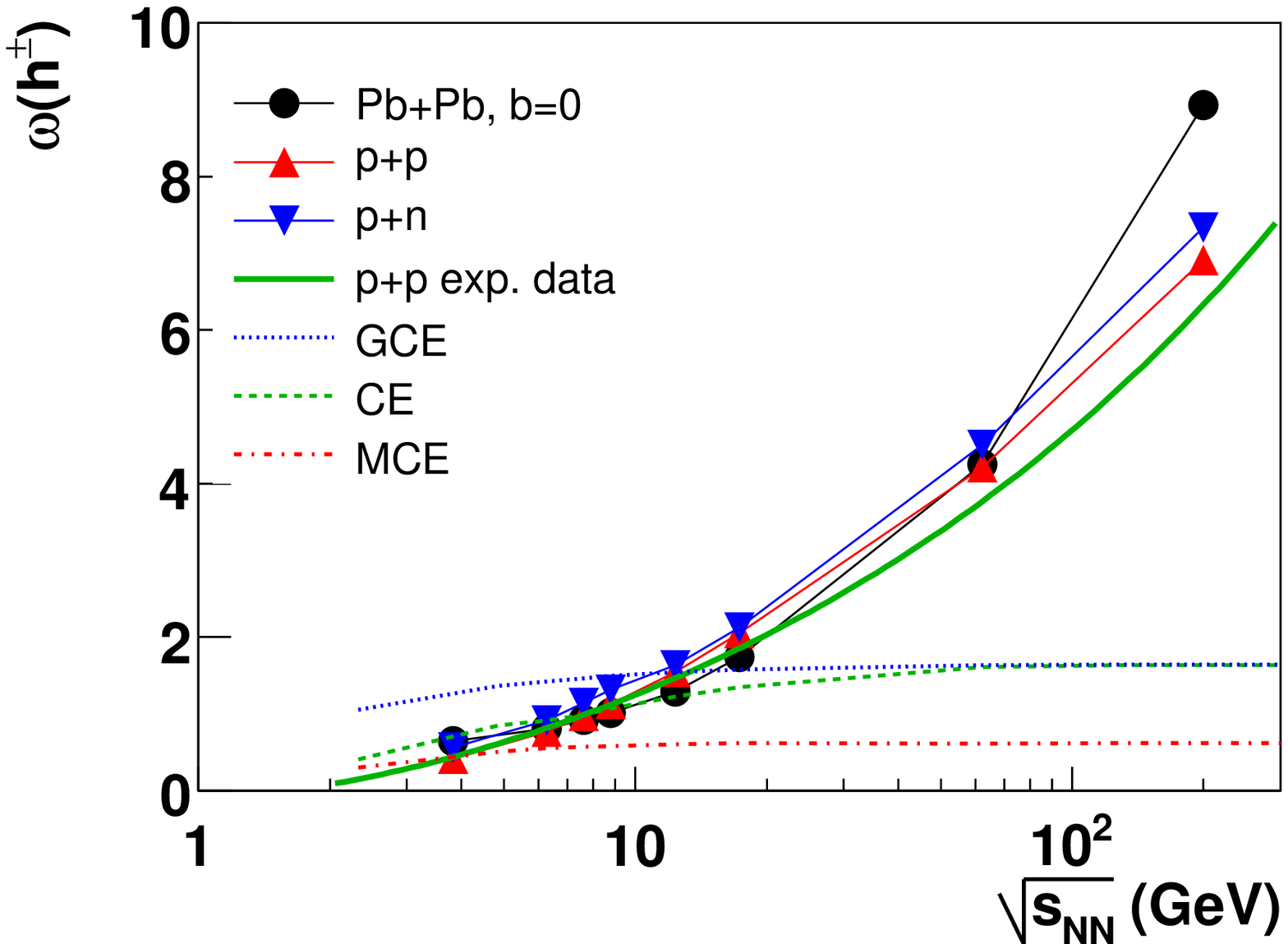}
\caption{\label{ed_w_urqmd}UrQMD results~\cite{Lungwitz:2007uc} of 
scaled variance of positively (left), negatively and all charged hadrons (right) in full phase space 
in inelastic p+p, p+n interactions and central Pb+Pb 
collisions as a function of collision energy 
in comparison to hadron-resonance gas model predictions~\cite{Begun:2006uu} 
for Pb+Pb collisions. For all charged hadrons the model predictions for p+p are
compared to a parametrization~\cite{Heiselberg:2000fk} of experimental data.}
\end{figure}

For p+p and p+n interactions all inelastic collisions are selected. For Pb+Pb the impact parameter of the collisions are set to $b=0$.
The calculations were performed for AGS, SPS and RHIC energies. 
The scaled variance predicted by the UrQMD model~\cite{Lungwitz:2007uc} 
is shown in figure~\ref{ed_w_urqmd}.

In the UrQMD model the multiplicity fluctuations are similar 
in nucleon-nucleon interactions and central heavy ion collisions.
Thus with respect to the scaled variance of multiplicity distributions UrQMD
behaves as a superposition model.
For positively and negatively charged hadrons the scaled variance is similar, 
whereas the values are about twice as high for all charged hadrons.

The energy dependence of the scaled variance is different from the predictions of 
the hadron-resonance gas model.
The scaled variance in the full phase space obtained by the UrQMD model increases monotonically with collision energy 
reaching values of up to $\omega=4.5$ for negatively charged hadrons at top RHIC energy.
The scaled variance predicted by the hadron-resonance gas model changes with energy in the AGS and SPS energy domain
but reaches its saturation value for energies of about $\sqrt{s_{NN}} \approx 100$~GeV.

In order to compare the UrQMD model to the experimental data, 
both the acceptance and the centrality selection of 
the NA49 experiment have to be taken into account.
The predictions of the model, published in \cite{Lungwitz:2007uc}, 
are compared to the experimental data in figures \ref{w_hp}-\ref{w_hpm}. 

Two different centrality selections (see section \ref{centsel})
are used in the model:
first, collisions with zero impact parameter (open symbols), 
second $1\%$ central collisions selected in the same way as it is done in the data 
using a simulation of the acceptance of the veto calorimeter (full symbols). 

The UrQMD model with collisions selected by their energy in the 
veto calorimeter is in agreement with data for all energies, acceptances and charges.
In the forward acceptance a UrQMD simulation for events with zero 
impact parameter ($b=0$) gives similar results, whereas the scaled variance for $b=0$ is 
smaller in the midrapidity and the full experimental regions, 
probably due to target participant fluctuations.

In the experimental data an increase of fluctuations 
is observed with decreasing rapidity (figure~\ref{ydep}).
The rapidity dependence of the scaled variance in data is reproduced by 
the UrQMD model with a similar centrality selection as in the data.

In the data an increase of scaled variance with decreasing 
transverse momentum is measured at forward rapidity (figure~\ref{ptdep}).
In the UrQMD model a similar trend is observed, but the scaled variance 
is underpredicted at very low transverse momenta,
which might be related to effects like Coulomb- and Bose-Einstein Correlations (HBT),
which are not implemented in the model.

The HSD transport approach, following a similar strategy as the UrQMD model,
yields similar results on scaled variance. The energy dependence of the scaled variance
for central ($b=0$) Pb+Pb collisions obtained by the HSD model are presented in 
\cite{Konchakovski:2007ss}. They are compared to the preliminary NA49 data on multiplicity
fluctuations~\cite{Lungwitz:2006cx} and both are in agreement in the forward acceptance.
Unfortunately HSD calculations for the larger acceptance used in this paper are not
available yet.

\subsection{Onset of Deconfinement}

The energy dependence of various observables shows anomalies at low 
SPS energies which might be related to the onset of 
deconfinement~\cite{Gazdzicki:1998vd,Gazdzicki:2004ef}. 
In \cite{Gazdzicki:2003bb} it is predicted that the onset of 
deconfinement should lead to additional fluctuations at medium SPS energies.
A "shark-fin" structure with a maximum near $80A$ GeV is predicted for the variable $R_e$ defined as:
\begin{equation}
R_e=\frac{(\delta S)^2 / S^2}{(\delta E)^2 / E^2} \,
\end{equation}
where $S$ is the entropy of the system and $E$ the energy of the collision 
which goes into produced particles (inelastic energy).
$R_e$ is approximately $0.6$ both in the hadron and quark gluon plasma phase, 
in the mixed phase it can reach values up to $0.8$.

In~\cite{Begun:2006uu} these fluctuations are used for an estimate of the behaviour of the multiplicity fluctuations
at the onset of deconfinement. In the mixed phase, the scaled variance of negatively charged particles is expected to
be increased by about $0.01$. This is smaller than the systematic 
error on the measurement of scaled variance, therefore the data can neither support
nor disprove the existence of a mixed phase at SPS energies.

\subsection{Critical Point}

In the phase diagram of strongly interacting matter it is expected 
that the hadron gas and quark-gluon-plasma regions are separated by a first
order phase transition line at high baryo-chemical potentials 
and lower temperatures. For higher temperatures and lower baryo-chemical potentials 
a cross-over between both phases is expected. 
The first order phase transition and the cross-over should be separated by the critical point.

If the freeze-out of the matter happens near the critical point, 
large fluctuations, for instance in multiplicity and transverse momentum, are
expected. In \cite{Stephanov:1999zu} it is estimated that 
the scaled variance in $4\pi$ should increase by at least $0.1$ near the critical point.
These critical fluctuations are expected to be located mainly at low transverse momenta.
Although the acceptance effect is unknown, no sign of the critical point 
is observed in the data on scaled variance.

\section{Summary}\label{c_summary}

The energy dependence of multiplicity fluctuations in central 
Pb+Pb collisions at $20A$, $30A$, $40A$, $80A$ and $158A$ GeV
was studied for positively, negatively and all charged hadrons. 
The full experimental acceptance ($0<y(\pi)<y_{beam}$) is divided 
into a midrapidity ($0<y(\pi)<1$) and a forward rapidity
region ($1<y(\pi)<y_{beam}$). 
At forward rapidity a suppression of fluctuations in comparison 
to a Poisson distribution is observed for positively and negatively charged hadrons.
At midrapidity and for all charged hadrons the fluctuations are higher.
Furthermore the rapidity and the transverse momentum dependence 
at $158A$ GeV 
was studied. The scaled variance increases for decreasing rapidity and
transverse momentum.

The results are in agreement with a UrQMD simulation using 
the same centrality selection as in the data and performed in the same acceptance.

The grand-canonical and canonical formulations of a hadron-resonance gas 
model~\cite{Begun:2006uu} are in disagreement with data.
They both overpredict fluctuations in the forward acceptance and predict a flat 
behaviour on rapidity and transverse momentum.
The micro-canonical formulation has a smaller scaled variance 
and can qualitatively reproduce the increase of fluctuations for low rapidities and
transverse momenta~\cite{Hauer:2007im}, but no quantitative 
calculations are available yet for the limited experimental acceptance.

The predicted maximum in fluctuations due to the phase transition from hadron-resonance 
gas to quark-gluon-plasma~\cite{Gazdzicki:2003bb} is expected to be smaller than the 
experimental errors and can therefore neither be confirmed nor disproved.

No sign of increased fluctuations as expected for a freeze-out near the 
critical point of strongly interacting matter was observed. 
The NA61 program~\cite{Gazdzicki:2006fy} will study both the energy 
and system size dependence of fluctuations in order to search for the critical point.

\begin{acknowledgments}
The author would like to thank M. Bleicher, M. Hauer, E. Bratkovskaya, V. Konchakovski, V. Begun, M. Gorenstein and I. Mishustin for 
fruitful discussions.\\
This work was supported by the US Department of Energy
Grant DE-FG03-97ER41020/A000,
the Bundesministerium fur Bildung und Forschung, Germany, 
the Virtual Institute VI-146 of Helmholtz Gemeinschaft, Germany,
the Polish State Committee for Scientific Research (1 P03B 006 30, 1 P03B 097 29, 1 PO3B 121 29, 1 P03B 127 30),
the Hungarian Scientific Research Foundation (T032648, T032293, T043514),
the Hungarian National Science Foundation, OTKA, (F034707),
the Polish-German Foundation, the Korea Science \& Engineering Foundation (R01-2005-000-10334-0),
the Bulgarian National Science Fund (Ph-09/05) and the Croatian Ministry of Science, Education and Sport (Project 098-0982887-2878).
\end{acknowledgments}

\bibliography{biblio} 

\end{document}